\documentclass[letterpaper,11pt,abstracton]{scrartcl}

\RequirePackage{fullpage}

\RequirePackage{enumerate}
\RequirePackage[latin1]{inputenc}
\RequirePackage{amsmath}
\RequirePackage{amssymb}

\usepackage{algorithmicx}
\usepackage{algorithm}
\usepackage[noend]{algpseudocode}

\usepackage{amsmath,amsfonts}
\usepackage{epsfig}

\newcommand{\qed}{\hfill$\diamond$}
\newcommand{\pf}{{\textbf {Proof: }}}
\newtheorem{theorem}{Theorem}[section]

\newtheorem{lemma}[theorem]{Lemma}
\newtheorem{tm}[theorem]{Theorem}
\newtheorem{claim}[theorem]{Claim}

\newtheorem{cor}[theorem]{Corollary}

\newtheorem{dfn}[theorem]{Definition}

\newtheorem{remark}[theorem]{Remark}

\begin{document}

\title{Recognizing Interval Bigraphs by Forbidden Patterns}

\date{}

\author{ Arash Rafiey \\ Indiana State University, IN, USA \\
arash.rafiey@indstate.edu \thanks{supported by NSF ( No. 1751765)}}

\maketitle

\begin{abstract}
Let $H$ be a connected bipartite graph with $n$ vertices and $m$ edges. We give an
$O(nm)$ time algorithm to decide whether $H$ is an interval bigraph.
The best known algorithm has time complexity $O(nm^6(m + n) \log n)$ and it was developed in 1997 \cite{muller}.
Our approach is based on an ordering characterization of interval bigraphs introduced by Hell and Huang \cite{hh2003}. We transform the problem of finding the desired ordering to choosing
strong components of a pair-digraph without creating conflicts. We make use of the structure of the pair-digraph as well as decomposition of
bigraph $H$ based on the special components of the pair-digraph. This way we make explicit what the difficult cases are and gain efficiency by isolating such situations.
\end{abstract}

\section{Introduction}


A bigraph $H$ is a bipartite graph with a fixed bipartition into
{\em black} and {\em white} vertices. We sometimes denote these
sets as $B$ and $W$, and view the vertex set of $H$ as partitioned
into $(B,W)$. The edge set of $H$ is denoted by $E(H)$. A bigraph
$H$ is called {\em interval bigraph} if there exists a family of
intervals (from real line) $I_v$ , $v \in B \cup W$, such that, for all $x \in B$
and $y \in W$, the vertices $x$ and $y$ are adjacent in $H$ if and
only if $I_x$ and $I_y$ intersect. The family of intervals is
called an {\em interval representation} of the bigraph $H$.

Interval bigraphs were introduced in \cite{hkm} and have been
studied in \cite{denver,hh2003,muller}. They are closely related
to interval digraphs introduced by Sen et. al. \cite{sdrw}, and in
particular, our algorithm can be used to recognize interval
digraphs (in time $O(mn)$) as well.


Recently interval bigraphs and interval digraphs became of interest in new areas
such as graph homomorphisms, cf. \cite{adjust}.

A bipartite graph whose complement
is a circular arc graph, is called a {\em co-circular arc bigraph}. It was shown in
\cite{hh2003} that the class of interval bigraphs is a subclass of co-circular arc bigraphs, corresponding
to those bigraphs whose complement is the intersection of a family of circular arcs no two of which
cover the circle. There is a linear time recognition algorithm for
co-circular arc bigraphs \cite{ross}. The class of interval bigraphs is a
super-class of proper interval bigraphs (bipartite permutation graphs) for which there is a linear time recognition algorithm 
\cite{hh2003, bss}.

Interval bigraphs can be recognized in polynomial time using the algorithm developed by Muller \cite{muller}. However, Muller's
algorithm runs in time $O(nm^6(n+m)\log n)$. This is in sharp contrast with the recognition of {\em interval  graphs}, for which
several linear time algorithms are known, e.g., \cite{booth,corneil,corneil09,habib,korte}.

In \cite{hh2003,muller} the authors attempted to give a forbidden structure
characterization of interval bigraphs, but fell short of the target. In this paper some light is shed on these attempts, as we clarify which situations
are not covered by the existing forbidden structures. We believe our algorithm can be used as a tool for producing the interval bigraph obstructions.
There are infinitely many obstructions and they do not fit into a few families of obstructions or at least we are not
able to describe them in such a manner. However, the main purpose of this paper is devising an efficient algorithm for recognizing interval bigraphs.

\begin{figure}[htbp] 
 \begin{center}
 \includegraphics[scale=0.5]{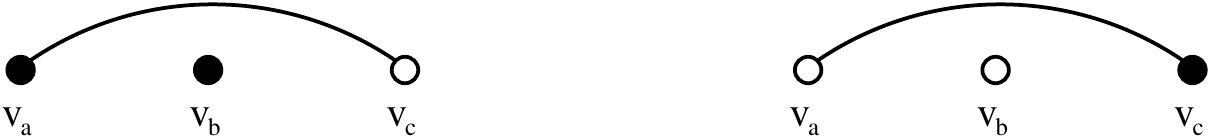}
 \caption{Forbidden Patterns}
 \label{fig:fig1}
 \end{center}
 \end{figure}

We use an ordering characterization of interval bigraphs introduced in \cite{hh2003}. A bigraph $H$ is interval 
if and only if its vertices admit a linear ordering $<$
without any of the forbidden patterns in Figure \ref{fig:fig1}. In such an ordering, if $v_a < v_b <
v_c$ (not necessarily consecutive) and $v_a,v_b$ have the same color and opposite to the color
of $v_c$ then $v_av_c \in E(H)$ implies that $v_bv_c \in E(H)$.

%
%
%

The vertex set of a graph $G$ is denoted by $V(G)$ and the edge
set of $G$ is denoted by $E(G)$. There are several
graph classes that can be characterized by existence of an ordering
without forbidden patterns. One such an example is the class of
interval graphs. A graph $G$ is an interval graph if and only if
there exists an ordering $<$ of the vertices of $G$ such that none
of the following patterns appears \cite{dama,dgr}.

\begin{itemize}
\item $v_a < v_b < v_c$, $v_av_c, v_bv_c \in E(G)$ and $v_av_b \not\in E(G)$
\item $v_a < v_b < v_c$, $v_av_c \in E(G)$ and $v_bv_c, v_av_b \not\in E(G)$
\end{itemize}

Some of the classes of graphs that have ordering characterizations without forbidden patterns are, proper 
interval graphs, comparability graphs, co-comparability graphs, chordal graphs, convex bipartite graphs, co-circular arc bigraphs,
proper interval bigraphs (bipartite permutation graph) \cite{esa-2014}. 

It is possible to view the ordering problem in some cases (e.g. interval bigraph, interval graph) 
as an instance of the 2-SAT problem together with transitivity clauses as follows. For every two vertices $u,v$ of $H$, we define 
a variable $X_{uv}$ which takes values zero and one only. We introduce clauses with 
two literals expressing the forbidden patterns, and also need to add transitivity clauses with three literals ($(X_{uv} \lor X_{vw} \lor X_{wu}) \land 
(X_{vu} \lor X_{wv} \lor X_{uw})$ ).  If $X_{uv}=1$ then we put $u$ before $v$; otherwise 
$v$ comes before $u$ in the ordering.  
However, we would like to consider a different approach proven to be more structural and successful in other ordering problems. 


\section{Basic definitions and properties}

We note that a bigraph is an interval bigraph if and only if each
connected component of it is an interval bigraph. In the remainder of
this paper, we shall assume that $H$ is a connected bigraph, with a
fixed bipartition $(B,W)$.

We define the following {\em pair-digraph} $H^+$ corresponding to the forbidden patterns in Figure \ref{fig:fig1}.
The vertex set of $H^+$ consists of pairs (vertices) $(u,v)$ with $u, v \in V(H)$, and $u \neq v$.

\begin{itemize}
\item There is an arc (in $H^+$) from $(u,v)$ to $(u',v)$ when $u,v$ have the same color and $uu' \in E(H)$ and $vu' \not\in E(H)$.

\item There is an arc (in $H^+$) from $(u,v)$ to $(u,v')$ when $u,v$ have different colors and $vv' \in E(H) $ and $uv \not\in E(G)$.
\end{itemize}

Note that if there is an arc from $(u,v)$ to $(u',v')$ then both $uv, u'v'$ are non-edges of $H$. 
For two vertices $\alpha,\beta \in V(H^+)$ we say $\alpha$ dominates $\beta$ or $\beta$ is dominated by 
$\alpha$ and we write $\alpha \rightarrow \beta$, if there exists an arc (directed edge) from $\alpha$ to $\beta$ in $H^+$. 
One should note that if $(x,y) \rightarrow (x',y')$ in $H^+$
then $(y',x') \rightarrow (y,x)$, so called {\em skew-symmetry} property. \\

\begin{lemma}\label{newpairsap}
Suppose $<$ is an ordering of $H$ without the forbidden patterns in Figure \ref{fig:fig1}.
If $u < v$ and $(u,v) \rightarrow (u',v')$ in $H^+$ then $u' < v'$. 
\end{lemma}
\pf Suppose $(u,v) \rightarrow (u',v')$. Now according to the definition of $H^+$; one of the following happens:
\begin{enumerate}
\item $u=u'$ and $u,v$ have different colors and $vv' \in E(H)$ and $uv \not\in E(H)$.
\item $v=v'$ and $u,v$ have the same color and $uu' \in E(H)$ and $vu' \not\in E(H)$.
\end{enumerate}

Suppose $u<v$ and (1) occurs. Since $uv \not\in E(H)$ and $vv'$ is an edge we must have $u<v'$ and hence $u' < v'$.

Suppose $u<v$ and (2) occurs. Since $u'u \in E(H)$  and $u'v \not\in E(H)$, we must have $u'<v$ and hence $u' < v'$.
\qed \\

In general, we shall write briefly {\em component} for {\em strong
component}. For a component $S$ of $H^+$, let $S' = \{(u,v):\ (v,u) \in S\}$ to be the {\em couple component} of $S$. A component in $H^+$ is called non-trivial if it contains more than one pair.

Note that the coupled components $S$
and $S'$ are either equal or disjoint; in the former case we say
that $S$ is a {\em self-coupled} component. Skew-symmetry of $H^+$ implies the following fact.

\begin{lemma} \label{couplesap}
If $S$ is a component of $H^+$ then so is $S'$.
\end{lemma}

\begin{dfn}[circuit]
A sequence $(x_0,x_1),(x_1,x_2),\dots, (x_{n-1},x_n),(x_n,x_0)$ of pairs in set $D \subseteq V(H^+)$ 
is called a {\em circuit} of $D$. 
\end{dfn}

\begin{lemma}\label{rainap}
If a component of $H^+$ contains a circuit then $H$ is not an interval bigraph. 
\end{lemma}
\pf Let $(x_0,x_1),(x_1,x_2),\dots, (x_{n-1},x_n),(x_n,x_0)$ be a circuit in component $S$ of $H^+$. By definition, there is a directed path $P_i$
from $(x_i,x_{i+1})$ to $(x_{i+1},x_{i+2})$, $0 \le i \le n$ (sum mod $n+1$) in $S$. Now by Lemma \ref{newpairsap} and following the vertices of $P_i$ 
when $x_i< x_{i+1}$ we conclude that $x_{i+1} < x_{i+2}$. 
Therefore, no linear ordering $<$ of $V(H)$ can have $x_i < x_{i+1}$ as otherwise we would have $x_0 < x_1 < \dots < x_n < x_0$.  
This would imply that we should have $x_0 > x_1$, $x_1 > x_2$, \dots, $x_{n} > x_0$, again not a linear ordering. \qed \\

If $H^+$ contains a self-coupled component then $H$ is not an interval bigraph. This is because a self-coupled component of $S$ contains both $(u,v)$ and $(v,u)$,
a circuit of length $2$ ($n=1$). As a remark we mention that if a component of $H^+$ contains a circuit then $H$ is not a co-circular arc bigraph \cite{esa-2012}.  \\

A tournament is a complete digraph with no directed cycle of length two and no self loop. A tournament is called transitive if it is acyclic, and does not 
contain a directed cycle. Now we have the following fact. 

\begin{lemma}\label{turnajap}
Suppose that $H^+$ contains no self-coupled components, and let $D$ be any
subset of $V(H^+)$ containing exactly one of each pair of coupled components.
Then $D$ is the set of arcs of a tournament on $V(H)$.
Moreover, such a $D$ can be chosen to be a transitive tournament if and only if $H$ is an interval bigraph.
\end{lemma}

In what follows, when we say a component we mean a non-trivial
component unless we specify otherwise. For simplicity, we shall also use $S$
to denote the sub-digraph of $H^+$ induced by $S$. \\

We shall say two edges $ab, cd $ of $H$ are {\em independent} if the subgraph
of $H$ induced by the vertices $a, b, c, d$ has just the two edges $ab, cd$. We shall say two disjoint induced subgraph $H_1,H_2$ of $H$ 
are {\em independent} if there is no edge of $H$ with one endpoint in $H_1$ and another endpoint in $H_2$. 
Note that if $ab, cd$ are independent edges in $H$ then the component of $H^+$
containing the pair $(a,c)$ also contains the pairs $(a,d), (b,c), (b,d)$. Moreover,
if $a$ and $c$ have the same color in $H$, the pairs $(a,c), (b,c), (b,d)$, $(a,d)$
form a directed four-cycle in $H^+$ in the given order; and if $a$ and $c$ have the
opposite color, the same vertices form a directed four-cycle in the reversed order.
In any event, an independent pair of edges yields at least four vertices in the
corresponding component of $H^+$. Conversely we have the following lemma.

\begin{lemma}\label{giantap}
Suppose $S$ is a component of $H^+$ containing a vertex
$(u,v)$. Then there exist two independent edges  $uu', vv'$ of
$H$, and hence $S$ contains at least the four vertices $(u,v),
(u,v'), $  \\ $(u',v), (u',v')$.
\end{lemma}
\pf Since $S$ is a component, $(u,v)$ dominates some pair of $S$ and is
dominated by some pair of $S$. First suppose $u$ and $v$ have
the same color in $H$. Then $(u,v)$ dominates some $(u',v) \in S$
and is dominated by some $(u,v') \in S$. Now $uu', vv'$ must be
edges of $H$ and $uv, uv', u'v, u'v'$ must be non-edges of $H$.
Thus $uu',vv'$ are independent edges in $H$. Now suppose $u$ and
$v$ have different colors. We note that $(u,v)$ dominates some
$(u,v') \in S$ and hence $uv$ is not an edge of $H$ and $vv'$ is
an edge of $H$. Since $(u,v')$ dominates some pair $(u',v') \in
S$, $uu'$ is an edge and $u'v'$ is not an edge of $H$. Now
$uu',vv'$ are edges of $H$ and $uv, uv', u'v, u'v'$ must be
non-edges of $H$. Thus $uu',vv'$ are independent edges in $H$. If $u,v$ have the same color then $S$ contains the directed cycle
$(u,v) \rightarrow (u',v) \rightarrow (u',v') 
\rightarrow (u',v) \rightarrow (u,v)$. If $u,v$ have different colors then $S$ contains the directed cycle 
$(u,v) \rightarrow (u,v') \rightarrow (u',v') 
\rightarrow (u',v) \rightarrow (u,v)$. 
\qed \\

Thus a component of $H^+$ must have at least four vertices. Recall that any pair $(u,v)$ in a component of $H^+$ must have
$u$ and $v$ non-adjacent in $H$.

\section{The Recognition algorithm}

We now present our algorithm for the recognition of interval
bigraphs. During the algorithm, we maintain a sub-digraph $D$ of
$H^+$. Initially, $D$ is empty; at successful termination, $D$
will be a transitive tournament as described in Lemma
\ref{turnajap}.

\begin{dfn}
Let $R$ be a subset of  $V(H^+)$. The {\em out-section} of
$R$, denoted by $N^+[R]$, consisting of all the pairs $(u,v)$ of $H^+$ such that either
$(u,v) \in R$ or $(u,v)$ is dominated by some $(u',v') \in R$. More formally $N^+[R] = \{ (u,v) | \ \  \exists (u',v') \in R \ \ s.t. \ \ (u',v') \rightarrow (u,v) \}$.  
\end{dfn}

In what follows for two sets $A,B$ the elements of $A$ that are not in $B$ are denoted by $A \setminus B$. 

We say a pair $(u,v)$ is {\em implied} by $R$ if $(u,v) \in N^+[R] \setminus R$.  

\begin{dfn}[Envelope]
Let $R$ be a subset of  $V(H^+)$. The {\em envelope} of $R$,
denoted by $N^*[R]$, is the smallest set of vertices that
contains $R$ and is closed under transitivity (if $(u,v),(v,w) \in N^*[R]$ then $(u,w) \in N^*[R]$) and out-section ( if $(u,v) \in N^*[R]$ and $(u,v) \rightarrow (u',v')$ in $H^+$ then $(u',v') \in N^*[R]$). 

\end{dfn}

{\textbf {Remark :}} For the purposes of the proofs we visualize taking the envelope of $R$
as divided into consecutive {\em levels}, where in zero-th level
we just replace $R$ by its out-section, and in each
subsequent level we replace $R$ by the out-section of its
transitive closure. 
The pairs in the envelope of $R$ can be
thought of as forming a digraph on $V(H)$, and each pair can be
thought of as having a label corresponding to its level. The pairs (arcs of the digraph) 
in $R$, and those implied by $R$ have the label $0$, arcs obtained
by transitivity from the arcs labeled $0$, as well as all arcs
implied by them have label $1$, and so on. More precisely $N^*[R]=R^0 \cup R^1 \cup \dots \cup R^k$. $R^0=N^+[R]$, level zero, and 
$R^i$, level $i \ge 1$ consisting of all the pairs in $R^{i-1}$ and 
the pairs $(u,v)$ where either $(u,v)$ is by transitivity over pairs $(u,u_1),(u_1,u_2),\dots,(u_{r-1},u_r),(u_r,v)$ in $R^{i-1}$ or 
there exists $(u',v') \rightarrow (u,v)$ where $(u',v')$ is by transitivity over pairs 
$(u',u'_1),(u'_1,u'_2),\dots,(u'_{r-1},u'_r),(u'_r,v')$ in $R^{i-1}$.

Note that $R \subseteq N^+[R] \subseteq N^*[R]$ and each of $R, N^+[R], N^*[R]$
may or may not contain a circuit.  For simplicity,  when $S$ is a component of $H^+$, let $N^+[S]$ and $N^*[S]$ denote $N^+[P], N^*[P]$ (respectively)  where $P= \{ (u,v) \in V(H^+) | (u,v) \in S \}$.   \\
Let $\mathcal{S}$ be a set of components of $H^+$. For simplicity, let $N^*[\mathcal{S}]$ denote $N^*[ \mathcal{P}]$ where $\mathcal{P}= \{(u,v) \in V(H^+) |  (u,v) \in S  \text { for some S in } \mathcal{S} \}$. 

The structure of components of $H^+$ is quite special,
and the trivial components interact in simple ways. A
trivial component will be called a {\em source component} if its
unique vertex has in-degree zero, and a {\em sink component} if
its unique vertex has out-degree zero. Before we describe the
structure, we establish a useful counterpart to Lemma
\ref{rainap}.

\begin{lemma} \label{oneap}
Let $S$ be a component, and $S'$ its coupled component. If both
$N^*[S]$ and $N^*[S']$ contain a circuit, then $H$ is
not an interval bigraph.
\end{lemma}
\pf It follows from the definition of $N^*[S]$ that if $N^*[S]$ contains a circuit then $V(S) \cap D =\emptyset $
 Therefore, $V(S' ) \subseteq D$ and hence, all the pairs in $N^*[S'] $ should be in $D$. This means
$D$ would contain a circuit, and hence, we get    
a contradiction by Lemma \ref{turnajap} because there is no total ordering. 

\qed

\begin{dfn}
Let $\mathcal{R}=\{ R_1 , R_2,...,R_k, S\}$ be a set of
components of $H^+$ such that $N^*[\mathcal{R}]$ contains a
circuit $C$. 
Let $W$ be an arbitrary subset of $\mathcal{R} \setminus \{S\}$ and let $W' = \{ R'_i \ \ | \ \ R_i \in W\}$.   
We say $S$ is a {\em dictator} for $C$ if the envelope of $W' \cup (\mathcal{R} \setminus W)$ 
also contains a circuit. In other words, by replacing some of the $R_i$'s with $R'_i$'s in $\mathcal{R}$ and taking 
the envelope we still get a circuit.
\end{dfn}

\begin{dfn}
A set $D_1 \subseteq V(H^+)$ is called {\em complete} if for every pair of coupled components 
$S,S'$ of $H^+$, exactly one of the $S \subseteq D_1$, $S' \subseteq D_1$ holds. 
\end{dfn}

Note that if the envelope of every complete set $D_1$ containing component $S$ has a circuit then $S$ is a dictator component. 


For the purpose of the algorithm once a pair $(x,y)$ is added into $D$ we assign a time (level) to $(x,y)$, that is the level in which $(x,y)$ is added into $D$. 
Each pair $(x,y)$ carries a dictator code, say $DCT(x,y)$; that shows the dictator component involved in creating a circuit containing $(x,y)$.  
%

\begin{dfn}
Suppose $D$ is a complete set. A pair $(x,y)$ of $H^+$ is called {\em original} 
\begin{itemize}
 \item if at least one of the $(x,y),(y,x)$ is not in $D$;
 \item if $(x',y') \rightarrow (x,y)$ then $(x',y') \in D$ is original;
 \item if $(x,y)$ is by transitivity over pairs $(x,w),(w,y) \in D$ then both $(x,w),(w,y)$ are original. 
 
\end{itemize}

\end{dfn}

During the computation of $N^*[D]$ we consider the circuits created by the original pairs. 
The purpose of introducing the original pairs is to detect all the
dictator components in one run of computing $N^*[D]$. 
%
%
%

In section \ref{circuit-structure} we show that if a circuit $C$ occurred by adding some 
pair into $D$ then its length is exactly $4$ and we
can identify a dictator component associated with $C$ by using $DCT(x,y)$, 
where $(x,y)$ is a pair of $C$. 
We show that $C$ is of form $C=(x_0,x_1),(x_1,x_2),(x_2,x_3),(x_3,x_0)$, where $x_0,x_3$ belong to the same color class while $x_1,x_2$ are contained 
in the opposite one; furthermore each pair $(x_i,x_{i+1})$, $0 \le i \le 3$ is an implied pair (not necessary from a component) or
inside a component and no pair $(x_i,x_{i+1})$ is by transitivity (the sum is taken module 3).\\

Another useful property is the following. Suppose $(x,w) \rightarrow (x,y)$, $(x,w') \rightarrow (x,y)$, and both $(x,w),(x,w')$ have
been added into $D$ at the same level. We show that $DCT(x,w)=DCT(x,w')$. 
 
%

\begin{dfn}
A pair $(x,y) \in D$ is {\em simple} if it belongs to $N^+[S]$ for
some component $S$. Otherwise we say $(x,y)$ is {\em complex}.

\end{dfn}

%

%
%

%


{\textbf {A  high overview of the algorithm :} }

Construct $H^+$ and consider its coupled components (recall that we mean strong components that are not trivial). 
If there is a component $S$ such that $S=S'$ then $H$ is not an interval bigraph. \\

In the first stage, we start with empty set $D$. Now from each pair 
of coupled components $S,S'$ we select one, say $S$. If $D \cup N^+[S]$ does not have a circuit then add $N^+[S]$ (all the pairs in $N^+[S]$) 
into $D$ and discard $N^+[S']$ from further consideration in this stage. Otherwise we discard $N^+[S]$ in this stage and add $N^+[S']$ into $D$ instead. 
If again $D$ has a circuit then we report $H$ is not an interval bigraph and exit. 
If we succeed in selecting exactly one of the coupled components $S,S'$ of $H^+$ then we proceed to the next stage. \\

In stage two, we compute $N^*[D]$ level by level. 
Suppose by adding a complex pair $(x_3,x_0)$ we encounter a circuit $C=(x_0,x_1),(x_1,x_2),(x_2,x_3),(x_3,x_0)$. Now we identify a dictator component $S_1$ that 
is responsible for $C$ ($S_1$ is obtained by tracing back the way of getting pair $(x_3,x_0)$). We add $S_1$ 
into set $\mathcal{DT}$ and continue computing $N^*[D]$. Note that we may encounter some other circuits while computing $N^*[D]$. As mentioned earlier,
we are looking for circuits with original pairs ($(x,y),(y,x)$) is not a circuit we are looking for).  \\

In stage three, we start with $D_1= \emptyset$ and for every $S_1 \in \mathcal{DT}$ we add $N^+[S'_1]$ into $D_1$ and discard $N^+[S_1]$. Moreover, 
for every (non-trivial strong) component $S_2 \in D \setminus \mathcal{DT}$ we add $N^+[S_2]$ into $D_1$ and discard $N^+[S'_2]$. 
Set $D=N^*[D_1]$. If there is a circuit in $D$ then report $H$ is not an interval bigraph and exit. Otherwise we proceed to the next stage. \\

In stage four, we add the remaining components (trivial strong components) of $H^+$ that are outside $D$, into $D$ one by one. 
At each step we add a sink component $S \subseteq V(H^+) \setminus D$ and discard $S'$ from further consideration.  \\

\begin{algorithm}
\caption{Algorithm for recognition of interval bigraphs}
  \label{alg-main}
  \begin{algorithmic}[1]
    \Function{IntervalBigraph}{$H$}
    \State {\textbf {Input:}} A connected bigraph $H$ with a bipartition $(B,W)$.
    \State {\textbf {Output:}} An ordering of the vertices of $H$ without patterns in Figure \ref{fig:fig1} or return 
    
    \hspace{2mm}  false.  
    
    \State Construct the pair-digraph $H^+$ of $H$, and compute its
components; if any are 

\hspace{2mm} self-coupled  report that $H$ is not an
interval bigraph.

\State Set $D$ to be an empty set. 
    
    \ForAll { coupled components $S,S' \subseteq V(H^+)$ } 
     
     \If { $D \cup N^+[S]$ does not have a circuit } 
      \State add $N^+[S]$ into $D$ and delete $N^+[S']$ from further consideration 
      
      \hspace{8mm}  in this step. \Comment { add $X$ to $D$  means add all the pairs of $X$ into $D$ }
      \ForAll { $(x,y) \in N^+[S]$ } set $DCT(x,y)=S$
      \EndFor 
     \Else {}  
     \If { $D \cup N^+[S'] $ does not have a circuit } 
       \State add $N^+[S']$ into $D$ and delete $N^+[S]$ from further consideration 
        
        \hspace{14mm} in this step.  
       \ForAll { $(x,y) \in N^+[S']$ } set $DCT(x,y)=S'$
      \EndFor
     \Else{} report that $H$ is not an interval bigraph. 
     \EndIf 
     
     \EndIf

    \EndFor 
  

    \State Set $En=N^*[D]$, and $\mathcal{DT}=\emptyset$  \Comment{ $\mathcal{DT}$ is a set of components} 
    
    \While { $\exists (x,y) \in En \setminus D$ } \Comment{we consider the pairs in $En$ level by level}
      
      \State Move $(x,y)$ into $D$ and set $DTC(x,y)=$ \Call{Dictator}{$x,y,D$} 
      \If { $D \cup \{(x,y)\}$ contains a circuit } 
       add $DCT(x,y)$ into $\mathcal{DT}$. 
      \EndIf
    \EndWhile

    \Comment{ $(x,y)$ is a complex pair }

      \State Let $D_1= \emptyset$. 
      \ForAll{components $S \in \mathcal{DT}$ }  
         add
$N^+[S']$ into $D_1$.
      \EndFor

   \ForAll { components  $R \in D \setminus\mathcal{DT}$ }
add $N^+[R]$ into $D_1$.

\EndFor

      \State Set $D= N^*[D_1]$. 
      \If {there is a circuit in $D$} 
      report $H$ is not an interval bigraph.
      \EndIf
 
     \While{ $\exists$ trivial component $S$ outside $D$, and $S$ is a sink component } 
      \State Add $S$ into $D$ and remove $S'$ from further consideration
 
    \EndWhile 
    
%
%
     \ForAll { $(u,v) \in D$ } set $u<v$, 
     
     \EndFor

     \State Let $v_1,v_2,\dots,v_n$ be the vertices of $H$ that respects the ordering $<$.

    \Comment{
     One can obtain the corresponding
interval representation of $H$ as described  in \cite{hh2003}.}

     \State Return $v_1,v_2,\dots,v_n$.

    \EndFunction
\Statex \Function{Dictator}{$x,y,D$}

 \If{$(x,y) \in N^+[S]$ for some component $S$ in $D$ } return $S$.
 
 \EndIf 
 
 \If {$x,y$ have different colors and $(u,y) \in D$  dominates $(x,y)$} 
 
   return $DCT(u,y)$. \Comment{ we mean the earliest pair $(u,y)$} 
 \EndIf
 
 \If {$x,y$ have the same color and $(x,w) \in D$ dominates $(x,y)$ } 
 
   return  $DCT(x,w)$
 
 \EndIf
 
 \If {$x,y$ have the same color and $(x,y)$ is by transitivity on
 
  $(x,w),(w,y) \in D$} return $DCT(w,y)$
 
 \EndIf
 
 \If {$x,y$ have different colors
and $(x,y)$ is by transitivity on

  $(x,w),(w,y) \in D$} return  $DCT(x,w)$
 \EndIf 
 
  \EndFunction 
    
  \end{algorithmic}

\end{algorithm}

\newpage

\section{Example :}

 \begin{figure}[htbp]
 \begin{center}
 \includegraphics[scale=0.8]{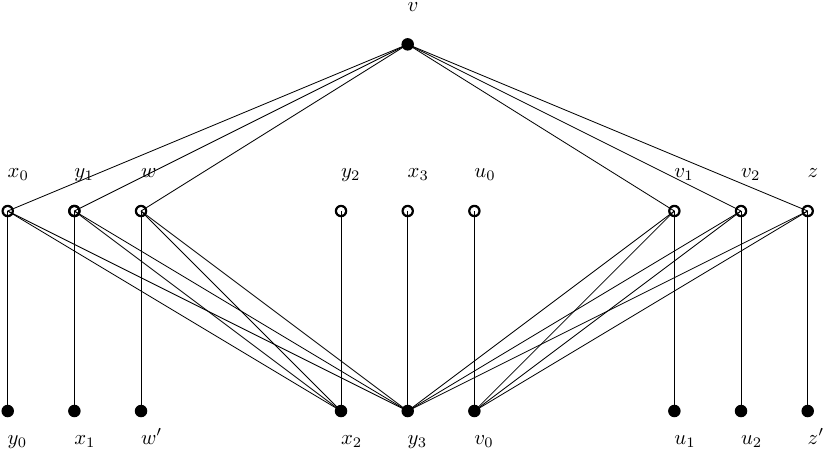}
 \caption{Bigraph $H$ is not an interval bigraph }
 \label{fig:Obst3}
 \end{center}
 \end{figure}

We apply the Algorithm 1 on the bigraph $H$ depicted in Figure \ref{fig:Obst3} and show that $H$ does not admit an 
ordering without the forbidden patterns in Figure \ref{fig:fig1}, and hence, $H$ is not interval. 

In fact we encounter a circuit after executing the lines 11-18 
and $N^*[D]$ also contains a circuit at line 22. Note that since $x_0y_0,x_1y_1,ww'$ are independent edges of $H$, $S_{x_0x_1},S_{x_1w}$ are components
of $H^+$. Since $u_1v_1,u_2v_2,z'z$ are independent edges, $S_{v_1u_2},S_{u_2z}$ are component of $H^+$. Finally, $x_2y_2,x_3y_3,v_0u_0$ are 
independent edges of $H$, and hence, $S_{x_2x_3},S_{x_3v0}$ are component of $H^+$ (recall that we mean non-trivial strong components).

Note that $(x_2,x_3),(x_3,v_0)$ are in the same
component since $x_2,y_3$ are adjacent to $w$ while
$v_0$ is not adjacent to $w$ and $y_3,v_0$ are adjacent to $v_1$
while $x_2v_1$ is not an edge of $H$, i.e. $(x_2,x_3) \rightarrow (x_2,y_3) \rightarrow (y_2,y_3) \rightarrow (y_2,v_1) \rightarrow (x_2,v_1) \rightarrow 
(x_2,v_0) \rightarrow  (w,v_0) \rightarrow  (w,u_0) \rightarrow (y_3,u_0) \rightarrow (y_3,v_0) \rightarrow (x_3,v_0)$. Therefore, $S_{x_2x_3}=S_{x_3v_0}$. 

Suppose we select components $S_{x_0x_1}$ and $S_{x_1w}$
and $S_{x_2x_3}$ and components $S_{u_1v_2},S_{v_2z'}$ at lines 5-11
of the algorithm and we add their pairs into $D$. This means $(x_0,x_1),(x_1,w),(x_1,x_2),(u_2,z),(x_3,v_0), \\ (v_1,u_2),(u_2,z)  \in D$. 
We have $(x_1,w) \rightarrow (x_1,x_2)$, 
$(u_2,z) \rightarrow (u_2,v)$, and $(x_3,v_0) \rightarrow
(x_3,v_1)$ in $H^+$. Therefore $(u_2,v),(x_3,v_1) \in N^+[D]$. 

Since the pairs $(v_1,u_2),(u_2,v)$ are in $N^+[D]$,
we have $(v_1,v) \in N^*[D]$ and consequently $(x_3,v_1),  \\ (v_1,v) \in N^*[D]$ implies that $(x_3,v) \in N^*[D]$. 
Now  $(x_0,x_1),(x_1,x_2),(x_3,v)$ are
placed in $N^*[D]$ ( at lines (5-11) of the algorithm).  
Moreover, $(x_3,v) \rightarrow (x_3,x_0) \in N^*[D]$, and hence, we have the circuit
$C = (x_0,x_1),(x_1,x_2),(x_2,x_3),(x_3,x_0)$ in $N^*[D]$. 

Note that since $y_3,v,v_0$  all are adjacent to $v_1,v_2,z$, selecting
$S_{v_2u_1}$ instead of $S_{u_1v_2}$ or selecting
$S_{z'v_2}$ instead of $S_{u_2z}$ would yield a circuit in
$N^*[D]$ as long as we select $S_{x_2x_3}$ to place in $D$. 
Moreover, selecting any two
components from $S_{x_0x_1},S_{x_1x_0},S_{x_1w},S_{wx_1}$ would also yield a
circuit in $D^*[D]$ as long as we select $S_{x_2x_3}$ at lines 5-12. 

Note that by adding $(x_3,v)$ into $N^*[D]$ we close circuit $C$. 
Now in order to obtain $DCT(x_3,x_0)$ we need to find $DTC(x_3,v)$. According to the rules of the algorithm, since $(x_3,v)$ is by transitivity 
on $(x_3,v_1), (v_1,v)$ and $x_3,v_1$ are white and $v$ is black, we have $DTC(x_3,v)=DTC(x_3,v_1)=S_{x_3v_0}=S_{x_2x_3}$ ( dictator component). 

Therefore, in order to avoid a circuit before line 12 of the algorithm we must
select $S_{x_3x_2}$. This means at line 20 we  place  $S_{x_3x_2}$ into $D_1$.  

Now if we select $S_{v_1u_2},S_{u_2z},S_{x_0x_1},S_{x_1w}$ at line 20 of the
algorithm we also place the following pairs  into $N^*[D_1]$. 
$(u_2,z) \rightarrow (u_2,v_0)$, $(x_3,x_2) \rightarrow
(x_3,x_0)$, $(x_0,x_1) \rightarrow (x_0,v)$, and $(v_0,x_3) \in S_{x_3x_2}$.

Therefore, by applying transitivity we would place $(x_0,v)$ into $N^*[D_1]$ (line 22)  and
now $(x_3,x_0),(x_0,v) \in N^*[D_1]$ would imply $(x_3,v) \rightarrow
(x_3,v_1)$. Therefore, we have the circuit
$(v_1,u_2), \\ (u_2,v_0),(v_0,x_3),(x_3,v_1)$ in $D$ (line 22). Selecting any 
two components from $S_{u_1v_2},S_{v_2u_1},S_{v_2z'},S_{z'v_2}$ instead
of $S_{u_1v_2},S_{v_2z'}$ would also yield a circuit. Therefore, $H$ is not an interval bigraph.

\section{Structural properties of the (strong) components of $H^+$}

\begin{lemma}\label{implyap}
A pair $(a,c)$ is implied by a component of $H^+$ if and only if
$H$ contains an induced path $a, b, c, d, e$, such that $N(a)
\subseteq N(c)$. If such a path exists, then the component $S$
implying $(a,c)$ contains all the pairs $(a,d), (a,e), (b,d),
(b,e)$.
\end{lemma}
\pf  If such a path exists, then $ab, de$ are independent edges
and so the pairs $(a,d), (a,e),$ \\ $(b,d), (b,e)$ lie in a
component by the remarks preceding Lemma \ref{giantap}. Moreover, $(a,d) \rightarrow (a,c)$ is in $H^+$; hence $(a,c)$ is
indeed implied by this component.

To prove the converse, suppose $(a,c)$ is implied by a component
$S$. We first observe that the colors of $a$ and $c$ must be the
same. Otherwise, say $a$ is black and $c$ is white, and there
exists a white vertex $u$ such that the pair $(u,c)$ is in $S$ and
dominates $(a,c)$. By Lemma \ref{giantap}, there would exist two
independent edges $uz, cy$. Looking at the edges and non-edges
amongst $u, c$ and $a, z, y$, we see that $H^+$ contains the arcs
$(u,c) \rightarrow (a,c) \rightarrow (a,y) \rightarrow (u,y).$
Since both $(u,c)$ and $(u,y)$ are in $S$, the pair $(a,c)$ must
also be in $S$, contrary to what we assumed.

Therefore, $a$ and $c$ must have the same color in $H$, say black.
In this case there exists a white vertex $d \in V(H)$ such 
that $(a,d) \in
S$ and $(a,d) \rightarrow (a,c)$. Hence $dc \in E(H)$ and $da \not\in E(H)$.
If there was also a vertex $t$ adjacent to
$a$ but not to $c$, then $at, cd$ would be independent edges of $H$, 
placing $(a,c)$ in $S$. Thus every neighbor of $a$ in $H$ is also
a neighbor of $c$ in $H$. Finally, since $(a,d) $ is in  
component $S$, Lemma \ref{giantap} yields vertices $b, e$ such
that $ab, de$ are independent edges in $H$. It follows that $a, b,
c, d, e$ is an induced path in $H$. \qed \\

We emphasize that $ab, de$ from the last Lemma are independent edges.
The inclusion $N(a) \subseteq N(c)$ implies the following corollary.

\begin{cor}\label{source-sinkap}
If there is an arc from a component $S$ of $H^+$ to a pair $(x,y)
\not\in S$ then $(x,y)$ forms a trivial component of $S$ which is a
sink component. If there is an arc to a component $S$ of $H^+$
from a pair $(x,y) \not\in S$ then $(x,y)$ forms a trivial component of
$H^+$ which is a source component. \qed
\end{cor}

In particular, we note that $H^+$ has no directed path joining two
components. To give even more structure to the components of
$H^+$, we recall the following definition. The {\em condensation}
of a digraph $D$ is a digraph obtained from $D$ by identifying the
vertices in each component and deleting loops and multiple edges.

\begin{lemma}\label{length_in_condensation}
Every directed path in the condensation of $H^+$ has at most three vertices.
\end{lemma}

\pf If a directed path $P$ in the condensation of $H^+$ goes
through a vertex corresponding to a  component $S$ in $H^+$, then
$P$ has at most three vertices by Corollary \ref{source-sinkap}.
Now suppose $P$ contains only vertices in trivial components and let 
$(x,y)$ be a vertex on $P$ which has both a predecessor and a
successor on $P$ otherwise we are done. First suppose $x$ and $y$ have the same color in $H$.  Then
the successor is some pair $(x',y)$ and the predecessor is some
pair $(x,y')$ and hence $xx', yy'$ are independent edges of $H$, and hence by Lemma \ref{giantap} 
$(x,y),(x',y),(x,y')$ belong to the same component of $H^+$, 
contradicting that $P$ goes through trivial components only.  Thus we continue by assuming that $x,y$ have the opposite color in $H$, and the
successor of $(x,y)$ in $P$ is some $(x,y')$ and the predecessor
is some $(x',y)$. Thus $xy$ is not an edge of $H$, whence
$x'y'$ must be an edge of $H$, otherwise we would have
independent edges $xx', yy'$ and conclude as above. By the same
reasoning, every vertex adjacent to $x$ is also adjacent to $y'$,
and every vertex adjacent to $y$ is also adjacent to $x'$. This
implies that $(x',y)$ has in-degree zero, and $(x,y')$ has
out-degree zero, and $P$ has only three vertices. \qed \\

\begin{figure}[htbp]
 \begin{center}
 \includegraphics[scale=0.5]{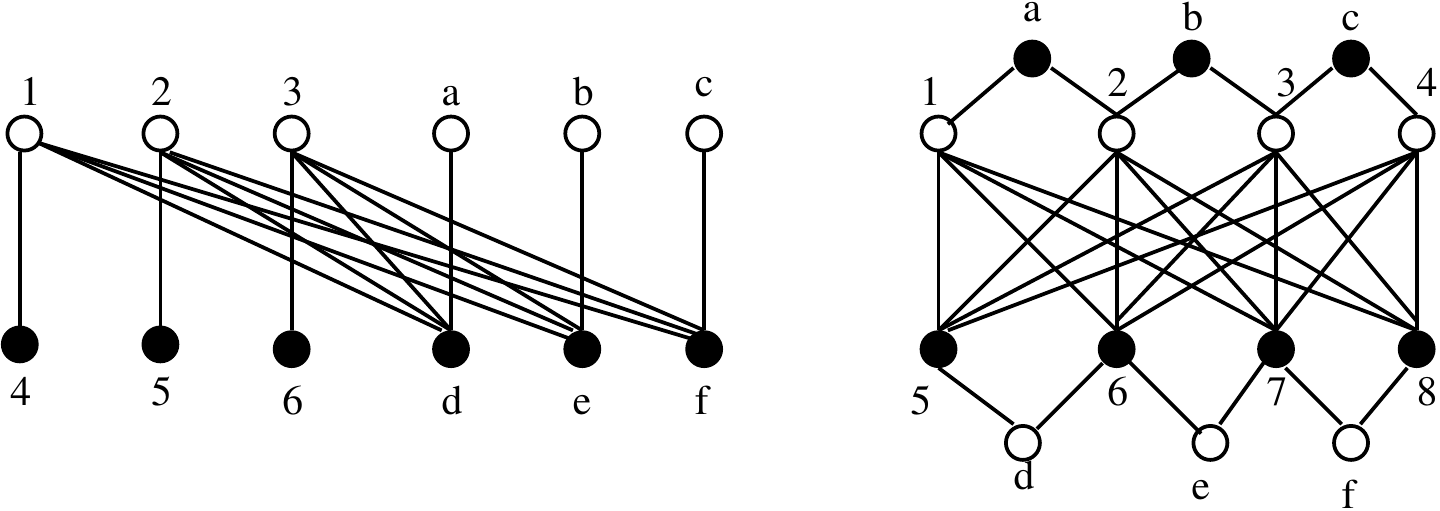}
 \caption{Exobicliques: 
 In left $B'=\{4,5,6,d,e,f\}$, $W'=\{1,2,3,a,b,c\}$ and $M=\{d,e,f\}$, $N=\{1,2,3\}$ 
 and $B'\setminus M=\{4,5,6\}$, $W' \setminus N=\{a,b,c\}$}
 
 \label{fig:fig2}
 \end{center}
 \end{figure}

\begin{dfn}

Let $H$ be a bigraph. We say an induced sub-bigraph $H'=(B',W')$ of $H$ is an {\em exobiclique} when the following hold. 
\begin{itemize}
 \item $B'$ contains a nonempty part $M$ and $W'$ contains a nonempty part 
$N$ such that $N \cup M$ induces a biclique in $H'$;
\item $B' \setminus M$ contains three vertices with
incomparable neighborhood in $N$ and $W' \setminus N$ contains three
vertices with incomparable neighborhoods in $M$ (see Figure \ref{fig:fig2}).

\end{itemize}
\end{dfn}

%

\begin{tm}\label{exohikoap}
If $H$ contains an exobiclique, as an induced subgraph, then $H$ is not an interval bigraph \cite{hh2003}.
\qed
\end{tm}

\begin{dfn} \label{pre-insect-def}
We say that a bigraph $H$ with bipartition $(B,W)$ is a {\em pre-insect},
if there exists disjoint induced subgraphs  
$H_1, H_2, \dots, H_k, X, Y, Z$ of $H$ where $k \geq 3$ and the
following properties are satisfied:

\begin{itemize}
\item[(1)] each $H_i$ is a component of $H' = H \setminus (X \cup Y \cup Z)$;

\item[(2)] $X$ is a complete bipartite graph;

\item [(3)] every vertex of $X$ is adjacent to all the vertices of opposite
color in $H'$;

\item [(4)] there are no edges between $Y$ and $H'$;

\item [(5)] there is no edge $ab$ in $Y$ such that both $a$ and $b$
are adjacent to all the vertices in $X$ of opposite color;

\item [(6)] if $Z$ is non-empty, then either

(i) every vertex of $Z$ is adjacent to all the vertices of opposite color in
each $H_i$ with $i  = 2,3$, or

(ii) every vertex of $Z$ is adjacent to at least one vertex of opposite
color in each $H_i$ with $i =2,3 $, and there are no edges between
$Z$ and $H_1$;

\item [(7)] every vertex of $Z$ is adjacent to all the vertices of opposite
color in $X \cup Z$.
\end{itemize}

\end{dfn}
We make the following observation on the components of a
pre-insect.

\begin{remark}\label{pre-insect_compap}
If $H$ is a pre-insect then all pairs $(u,v)$ where $u \in H_i$ and $v
\in H_j$, for some fixed $i \neq j$, are contained in the same
component $S^{(i,j)}$ of $H^+$. Moreover, if $Z$ is not empty then we 
have $S^{(1,2)}=S^{(1,3)}=\dots=S^{(1,k)}$, and if $Z$ is empty then $(i,j) \neq
(i',j')$ implies that $S^{(i,j)},S^{(i',j')}$ are distinct
components of $H^+$.
\end{remark}

In the sequel, we shall use $S_{uv}$ to denote the component of $H^+$
containing the pair $(u,v)$. Thus $S_{uv}$ and $S_{vu}$ are coupled
components of $H^+$. \\

We shall say that a vertex $v$ is {\em completely adjacent} to a subgraph
$V$ of $H$ if $v$ is adjacent to every vertex of opposite color in $V$.
We shall also say that $v$ is {\em completely non-adjacent} to $V$ if it
has no edges to $V$.

\begin{tm}\label{pre-exoap}
Suppose that $H^+$ has no self-coupled components.

Suppose $H$ has three vertices $u, v, w$ such that $S_{uv}, S_{vw}$ are
components of $H^+$ and $S_{uv} \neq S_{vw}$ and $S_{uv} \neq S_{wv}$. 
Then $H$ is a pre-insect and $u, v, w$ belong to different
connected components of $H'$ (see definition \ref{pre-insect-def}). 

Moreover, in this case $S_{wu} \neq S_{uv}$ and $S_{wu} \neq S_{vw}$.
If all $S_{uv}, S_{vu}, S_{vw}, S_{wv}, S_{uw}, S_{wu}$ are pairwise distinct
then the subgraph $Z$ is empty; otherwise $Z$ is non-empty and either
$S_{uw} = S_{uv}$ or $S_{uw} = S_{vw}$.
\end{tm}

\pf
First we observe that the skew-symmetry of $H^+$ implies that
$S_{uv}  \neq S_{vw}$ and $ S_{uv} \neq S_{wv}$. It is also implies that 
$S_{vu} \neq S_{wv}$ and $ S_{vu} \neq S_{vw}$. So we may freely use
any of these properties in the proof.

Since $S_{uv}, S_{vw}$ are  components of $H^+$, by Lemma \ref{giantap}, there exit $u',v',v'',w' \in V(H)$ such that 
$uu', vv' $ are two independent edges of $H$ and $vv'', ww'$ are two independent edges of $H$. Assume that $u, v, w$ are of the
same color - in case when $u, v$ are of different colors, we
switch the names of $u, u'$ and when $v, w$ are of different
colors, we switch the names of $w, w'$.

Since $H^+$ has no self-coupled components,  
$S_{uv},S_{vu}, S_{vw}, S_{wv}$ are pairwise distinct components of $H^+$.
Hence by Corollary \ref{source-sinkap} there is no directed path
in $H^+$ between any two of them.

We claim that $uu', ww'$ are independent edges. Indeed, an
adjacency between $u$ and $w'$ in $H$ would mean an arc from
$(u,v)$ to $(w',v)$ in $H^+$ and an adjacency between $u'$ and $w$
in $H$ would mean a directed edge from $(w,v)$ to $(u',v)$, both
contradicting our assumptions. It follows that $S_{uw}$ and
$S_{wu}$ are also  components.

If both $uv''$ and $wv'$ are edges of $H$ then there is an arc
from $(u,v')$ to $(u,w)$, implying $S_{uv}=S_{uw}$ and there is an arc from $(v'',w')$ to $(u,w')$ 
implying that $S_{vw}=S_{uw}$, and hence $S_{uv}=S_{vw}$, a contradiction.
So either $uv''$ or $wv'$ is not
an edge of $H$. By symmetry, we may assume that $wv'$ is not an edge
of $H$. Hence $uu', vv'$, and $ww'$ are three pairwise independent edges
of $H$.

Let $H_1,H_2,H_3$ be three maximal connected induced subgraph of $H$ containing $uu', vv', ww'$ respectively which are disjoint and independent. 

Let $X$ be the set of vertices completely adjacent to $H_1 \cup H_2 \cup H_3$.
Let $Y_1$ be the set of vertices completely non-adjacent to $H_1 \cup H_2 \cup H_3$,
and let $H_4,H_5,\dots,H_k$ be the connected component of $Y_1$ such that each $H_i$, $4 \le i \le k$, is completely adjacent to $X$. 

We shall also use $X, Y_1$, etc., to denote the subgraphs of $H$ induced by these
vertex sets.

We let $H'$ consist of $H_1, H_2, H_3, H_4, \dots, H_k$. 
We also let $Y = Y_1 \setminus
H'$, and let $Z = H\setminus (H' \cup X \cup Y)$. We now verify
the conditions (1-7).

It follows from the definition that every vertex of $X$ is completely adjacent
to $H'$, every vertex of $Y$ is completely non-adjacent to $H'$, and every
vertex of $Z$ has neighbors from at least two of $H_1, H_2, H_3$ (but is
not completely adjacent to $H_1 \cup H_2 \cup H_3$).

We claim that $X$ is a complete bigraph. Indeed, suppose that $x, x'$ are
vertices of $X$ of opposite colors, where $x$ is of the same color as $u$.
If $x, x'$ are not adjacent then $(u',v), (u',x'), (x,x'), (x,v), \\ (w',v)$ is a directed
path in $H^+$ from $S_{uv}$ to $S_{wv}$, a contradiction.

The definition of $H'$ also implies that if $yy'$ is an edge of $Y$ then $y, y'$
cannot both be completely adjacent to $X$.

Let $a \in H_1, b \in H_2, c \in H_3$ be three vertices of the same color.
Suppose that some $z \in V(H)$ is adjacent to two of these vertices but not to the
third one; say, $z$ is adjacent to $b$ and $c$ but not to $a$. Clearly, $z \in Z$.
Let $a'$ be any vertex in $H_1$ adjacent to $a$. Then $(a',b), (a',z), (a,z), (a,c)$
is a directed path from $S_{uv}$ to $S_{uw}$, implying $S_{uv} = S_{uw}$.
This property implies that if $S_{uv}, S_{vu}, S_{vw}, S_{wv}, S_{uw}$, and
$S_{wu}$ are pairwise distinct then $Z$ is empty. (The converse is also true,
i.e., if $Z$ is empty then $S_{uv}, S_{vu}, S_{vw}, S_{wv}, S_{uw}$, and
$S_{wu}$ are pairwise distinct.)

Since $S_{uv} \neq S_{vw}, S_{wv}$, the same property implies that every vertex
of $Z$ adjacent to vertices in $H_1$ and in $H_3$ must be completely adjacent
either to $H_1 \cup H_2$ or to $H_2 \cup H_3$.

If some vertex of $z\in Z$ is completely adjacent to $H_2 \cup H_3$ then
$z$ is not completely adjacent to $H_1$ and hence the above property implies
that $S_{uv} = S_{uw}$. Similarly, if some vertex of $Z$ is completely adjacent to
$H_1 \cup H_2$ then we have $S_{wu} = S_{wv}$ (i.e., $S_{uw} = S_{vw}$).
Since $S_{uv} \neq S_{vw}$, $Z$ cannot contain both a vertex completely
adjacent to $H_1 \cup H_2$ and a vertex completely adjacent to $H_2 \cup H_3$.
Therefore, when $Z$ is not empty, either $H_1$ or $H_3$ enjoys
a "special position", in the sense that

\begin{itemize}
\item
each vertex of $Z$ is adjacent to at least one vertex in $H_2$ and at least
one vertex in $H_3$ and is nonadjacent to at least one vertex in $H_1$.
Moreover, if it is also adjacent to a vertex in $H_1$, then it is 
completely adjacent to $H_2 \cup H_3$.
(This corresponds to the case $S_{uw} = S_{uv}$.)

\item
each vertex of $Z$ is adjacent to at least one vertex in $H_1$ and at least
one vertex in $H_2$ and is nonadjacent to at least one vertex in $H_3$.
Moreover, if it is also adjacent to a vertex in $H_3$ then it is
completely adjacent to $H_1 \cup H_2$.
(This corresponds to the case $S_{uw} = S_{vw}$.)
\end{itemize}

In either case, we have $S_{wu} \neq S_{uv}$ and $ S_{wu} \neq S_{vw}$. 
Note that by symmetry the two above are corresponding to condition (6) of definition of pre-insect.

Finally, we show that every vertex of $Z$ is completely adjacent to $X \cup Z$.
Let $z \in Z$. From above we know that either $z$ has neighbors in $H_1$ and
in $H_2$, or $z$ has neighbors in $H_2$ and in $H_3$. Assume that $aa' \in E(H_1)$
and $bb' \in E(H_2)$ where $a',b'$ are neighbors of $z$. (A similar argument applies in the other
case.) Suppose that $z$ is not adjacent to a vertex $x' \in X$ of the opposite color.
Since
each vertex of $X$ is completely adjacent to $H_1 \cup H_2$, the vertex $x'$ is
adjacent to both $a$ and $b$. Thus $z,a',a,x',b,b',z$ is an induced 6-cycle in $H$,
which is easily seen to imply that $S_{uv} = S_{vu}$, a contradiction. Suppose
now that $z$ is not adjacent to a vertex $z' \in Z$ of opposite color. Then as
above $z'$ has neighbors $a_1 \in H_1$ and $b_1 \in H_2$. Choose $a_1,b_1$ so that
 $a_1, a'$ have the minimum distance in $H_1$ and
$b_1, b'$ have the minimum distance in $H_2$. It is easy to see that there is
an induced cycle of length at least six in $H$, using vertices $z, a_1, a', b_1, b'$
a shortest path in $H_1$ joining $a_1, a'$ and a shortest path in $H_2$ joining
$b_1, b'$. This implies again that $S_{uv} = S_{vu}$, a contradiction. \\
\qed

We now consider the possibility that for three vertices $u,
v, w$ of $H$, the components $S_{uv}, S_{vw}$ coincide; of course
then this common component $S_{uv} = S_{vw}$ is a  component.

\begin{lemma}\label{transitiveap}
Suppose that $H^+$ has no self-coupled components.
If for three vertices $u, v, w$ of $H$, $S_{uv},S_{vw}$ are component of $H^+$ and 
$S_{uv} = S_{vw}$ then we also have $S_{uv} = S_{uw}$. 
\end{lemma}

\pf Since $S_{uv}$ is a  component, there are independent edges
$uu', vv'$;  similarly, there are independent edges $vv'', ww'$. We
may assume that $u, v, w$ are of the same color - in case when $u,
v$ are of different colors, we switch the names of $u, u'$ and
similarly for $v, w$.

We claim that neither $uw'$ nor $wu'$ is an edge of $H$. Indeed, if $uw'$ is
an edge of $H$ then $uw', vv'$ are independent edges of $H$, which implies
that $S_{uv} = S_{w'v} = S_{wv}$. However, we know by assumption
$S_{uv} = S_{vw}$. Thus $S_{wv} = S_{vw}$, a contradiction. Similarly, if
$wu'$ is an edge then $wu', vv''$ are independent edges, which implies
$S_{vw} = S_{vu'} = S_{vu}$. Since $S_{uv} = S_{vw}$, we have $S_{uv} = S_{vu}$,
a contradiction.

If $uv''$ and $wv'$ are both edges of $H$ then they are independent and we have
$S_{uv} = S_{uv'} = S_{uw}$. By symmetry, we may assume that $wv'$ is not an edge
of $H$. Hence we obtain three pairwise independent edges $uu', vv', ww'$ of $H$.

Following the proof of Theorem \ref{pre-exoap}, we define the subgraphs $H', X, Y, Z$.
Since $S_{uv} = S_{vw}$, the set $Z$ is not empty. Each vertex of $Z$ has
neighbors in at least two of $H_1, H_2, H_3$ but is not completely adjacent to
$H_1 \cup H_2 \cup H_3$. It is not possible that some vertex of $Z$ is adjacent
to vertices in $H_1$ and in $H_3$ but nonadjacent to a vertex in $H_2$, as
otherwise we would have $S_{uv} = S_{wv}$ and  because $S_{uv} = S_{vw}$, we have  
$S_{wv} = S_{vw}$, a contradiction. If some vertex of $Z$ adjacent to vertices
in $H_2$ and in $H_3$ but nonadjacent to a vertex in $H_1$ then $S_{uv} = S_{uw}$;
similarly, if some vertex adjacent to vertices in $H_1$ and in $H_2$ but nonadjacent
to a vertex in $H_3$ then $S_{uw} = S_{vw}$. This completes the proof.
\qed

We now summarize the possible structure of the six related components
$S_{uv}, S_{vu}, S_{vw}, S_{wv}, S_{uw}$, and $S_{wu}$. Theorem \ref{pre-exoap}
and Lemma \ref{transitiveap} imply the following corollary.

\begin{cor}\label{jing3.2(C)ap}
Suppose that $H^+$ has no self-coupled components.

Let $u, v, w$ be three vertices of $H$ such that $S_{uv}$ and
$S_{vw}$ are  components of $H^+$. Then $S_{uw}$ is also a
 component of $H^+$.

Moreover, one of the following occurs, up to a permutation of $u, v, w$.

\begin{enumerate}
\item [(i)]
$S_{uv}, S_{vu}, S_{vw}, S_{wv}, S_{uw}$, and $S_{wu}$ are pairwise distinct;

\item [(ii)]
$S_{uv} = S_{uw}$, $S_{wu} = S_{vu}$, $S_{vw}$, $S_{wv}$ are
pairwise distinct;

\item [(iii)]
$S_{uv} = S_{vw} = S_{uw}$ and $S_{vu} = S_{wv} = S_{wu}$ are distinct.
\end{enumerate}
\qed
\end{cor}

\section{Correctness of lines 4--12}

We consider what happens when a circuit is formed
during the execution of lines 4--12 of our algorithm; our goal is to prove that
in such a case $H$ contains an exobiclique and hence is not an interval
bigraph. Note that we only get to lines 4--12 if $H^+$ has no self-coupled
components, so we do not need to explicitly make this assumption.

\begin{lemma}\label{new}
Let $S_1$ and $S_2$ be two components in $D$ and $D$ does not have
a circuit. Suppose $(y,y') \in N^+[S_1]$, and $(z,z') \in N^+[S_2]$
where $y,y'$ have the same color and $yz',y'z$ are edges of $H$.
Then $yz,y'z'$ are edges of $H$.
\end{lemma}
\pf For contradiction suppose $y'z'$ is not an edge of $H$. Now
$(z,z') \rightarrow (y',z') \rightarrow (y',y)$. Now by skew
symmetry property there exist $(w,w') \in S'_1$ such that $(y',y)
\rightarrow (w,w')$ and by definition of $N^+[S_2]$ there exists
$(v,v') \in S_2$ such that $(v,v') \rightarrow (z,z')$. Thus there
exists a path in $H^+$ from  vertex $(v,v')$ in $S_2$ to vertex
$(w,w')$ in $S'_1$. By Corollary \ref{source-sinkap} and Lemma
\ref{length_in_condensation} this happens only if $S_2=S'_1$. But
this is a contradiction since it would imply that both $S_1$ and
$S'_1$ are in $D$. By similar argument $zy$ is an edge of $H$.
\qed

\begin{tm}\label{trichotomyap}
Suppose that within lines 4--12 we have so far constructed a $D$
without circuits, and then for the next  component $S$ we find
that $D \cup N^+[S]$ has circuits. Let 
$C:(x_0,x_1),(x_1,x_2),\dots, (x_n,x_0)$ be a shortest circuit in $D
\cup N^+[S]$. Then one of the following must occur.

\begin{itemize}

\item [(i)] $H$ is a pre-insect with empty $Z$, each $x_i$ belongs to some
subgraph $H_{a_i}$, and $i \neq j$ implies $a_i \neq a_j$, or

\item[(ii)] $H$ is a pre-insect with non-empty $Z$, each $x_i$ belongs to some
subgraph $H_{a_i}$ with $i > 1$, and $i \neq j$ implies $a_i \neq a_j$, or

\item[(iii)] $H$ contains an exobiclique.

\end{itemize}

\end{tm}

\pf From the way the algorithm constructs $D$, we know that each
pair $(x_i,x_{i+1})$ either belongs to or is implied by a
component in $D \cup N^+[S]$. The length of $C$ is at least three,
i.e., $n \geq 2$, otherwise $S_{x_0x_1}$ and $S_{x_1x_0}$ are both
in $D \cup N^+[S]$, contrary to our algorithm.

We first show that no two consecutive pairs of $C$ are both implied by
 components. Indeed, suppose that for some subscript $s$, both
$(x_{s-2},x_{s-1})$ and $(x_{s-1},x_s)$ are implied by
components. Then by Lemma \ref{implyap}, there are induced paths
$x_{s-2}, x, x_{s-1}, y, z$ and $x_{s-1}, u, x_s, v, w$ with
$N(x_{s-2}) \subseteq N(x_{s-1}) \subseteq N(x_s)$. Since $x, y$ are
adjacent to $x_{s-1}$, they are adjacent also to $x_s$. Thus
$x_{s-2}, x, x_s, y, z$ is an induced path in $H$ (with
$N(x_{s-2}) \subseteq N(x_s)$). By Lemma \ref{implyap},
$(x_{s-2},x_s)$ is implied by $S_{yx_{s-2}}$. We know that
$S_{yx_{s-2}}$ is in $D \cup N^+[S]$ because it implies
$(x_{s-2},x_{s-1})$. Hence $(x_{s-2},x_s)$ is also in $D \cup
N^+[S]$. Replacing $(x_{s-2},x_{s-1}), (x_{s-1},x_s)$ with
$(x_{s-2},x_s)$ in $C$, we obtain a circuit in $D \cup N^+[S]$
shorter than $C$, a contradiction.

Suppose that for some $s$ both $(x_{s-2},x_{s-1})$ and
$(x_{s-1},x_s)$ belong to  components. By Lemma
\ref{transitiveap}, $(x_s,x_{s-2})$ also belongs to a component.
Consider $S_{x_{s-2}x_{s-1}}, S_{x_{s-1}x_s}$, and
$S_{x_sx_{s-2}}$. Suppose that any two of these are equal. Then
they are equal to the component coupled with the third one, by
Lemma \ref{transitiveap}. This means that either
$(x_{s-1},x_{s-2})$, or $(x_s,x_{s-1})$, or $(x_{s-2},x_s)$ is
contained in $D \cup N^+[S]$, each resulting in a shorter circuit,
and a contradiction. Therefore, by Corollary \ref{jing3.2(C)ap},
we have the following cases:

\begin{itemize}
\item [(1)]
the six components $S_{x_{s-2}x_{s-1}}, S_{x_{s-1}x_{s-2}}, S_{x_{s-1}x_s},
S_{x_sx_{s-1}}, S_{x_sx_{s-2}}, S_{x_{s-2}x_s}$ are pairwise distinct;

\item [(2)]
$S_{x_{s-2}x_{s-1}} = S_{x_{s-2}x_s}$;

\item [(3)]
$S_{x_{s-2}x_{s-1}} = S_{x_sx_{s-1}}$; or

\item [(4)]
$S_{x_{s-1}x_s} = S_{x_{s-2}x_s}$.
\end{itemize}

Since (2), (3), and (4) result in a circuit in $D \cup N^+[S]$ shorter than $C$,
we must have (1).
By Theorem \ref{pre-exoap}, $H$ is a pre-insect with empty
set $Z$. So  either  each $x_i$ is in $H'$, implying the case (i), or
some $x_j$ belongs to $X \cup Y$. As in the proof of Theorem \ref{pre-exoap},
let $H_1, H_2, H_3, \dots $ be the connected components of $H'$ where
$x_{s-2} \in H_1, x_{s-1} \in H_2, x_s \in H_3$. Without loss of generality
assume that $x_{s+1}, \dots, x_{t-1} \in X \cup Y$ and $x_t \in H_d$. Note
that $d \neq 3$, by the minimality of $C$.

We show that $S_{x_{t-1}x_t}$ is a trivial component. Otherwise,
by Lemma \ref{giantap}, we obtain two independent edges $x_{t-1}u$
and $x_tv$. It is easy to see that $x_{t-1}u$ lies in $Y$ and the
vertex $v$ is either in $H_d$ or in $X$. We assume that $x_t$ is
of the same color as $x_{t-1}$ (the discussion is similar when
they are of different colors). We know from above that either
$x_{t-1}$ or $u$ is not adjacent to some vertex in $X$ of opposite
color. Assume first that $x_{t-1}$ is not adjacent to $w \in X$ of
opposite color. Since each vertex of $X$ is completely adjacent to
$H'$, $w$ is adjacent to $x_t$ and a vertex $w' \in H_3$ (note
that $H_3$ contains $x_s$). We see now that $x_{t-1}u$ is
independent with both $wx_{t}$ and $ww'$, which means that
$S_{x_{t-1}x_t} = S_{x_{t-1}x_s}$. We have a shorter circuit
$(x_s,x_{s+1}), \dots, (x_{t-1},x_s)$, a contradiction. The proof
is similar if $u$ is not adjacent to some vertex in $X$. So
$S_{x_{t-1}x_t}$ is a trivial component, and hence $(x_{t-1},x_t)$
is implied by some  component.

By Lemma \ref{implyap} there is an induced path $x_{t-1},y,x_t,z,w$ in
$H$ such that $N(x_{t-1}) \subseteq N(x_t)$, which implies that $y
\in X$ and $x_{t-1} \in Y$. Clearly, $w \notin X \cup H_d$ as it
is not adjacent to $y \in X$ and $z \notin Y$ as it is adjacent to
$x_t$. It follows that $z \in X$ and $w$ is in $Y$.  Note that
$(x_{t-1},z)$ is in a  component. Now $(x_{t-1},z) \rightarrow
(x_{t-1},v)$ for some $v \in H_2$. If $v, x_{s-1}$ have the same
color and in this case $(x_{t-1},z) \rightarrow (x_{t-1},x_{s-1})$
and hence we a get a shorter circuit. If $v,x_{s-1}$ have
different colors then there is also circuit
$(x_0,x_1),...,(x_{s-2},v),(v,x_{s}),(x_s,x_{s+1}),...,(x_n,x_0)$
in $D$ since $(x_{s-2},v),(x_{s-2},x_{s-1})$ are in the same
component, and $(v,x_s),(x_{s-1},x_{s})$ are in the same component.
Therefore, we get a shorter circuit.

It remains to consider the situation where consecutive pairs of
$C$ always alternate, in belonging to, and being implied by, a
component. Suppose that $(x_i,x_{i+1})$ is implied by a
 component. By Lemma \ref{implyap}, there is an induced
path $x_i,a,x_{i+1},b,c$ with $N(x_i) \subseteq N(x_{i+1})$. Note that
$x_i$ and $x_{i+1}$ have the same color.

We show that $x_{i+2}$ has color different from that of $x_i$. For
a contradiction, suppose that they are of the same color. Let
$x_{i+1}f, x_{i+2}g$ be independent edges in $H$; such edges exist
because $(x_{i+1},x_{i+2})$ belongs to a component. Since $N(x_i)
\subseteq N(x_{i+1})$ and $x_{i+1}g$ is not an edge of $H$, we conclude that
$x_ig$ is not an edge of $H$. We also see that $bx_{i+2}$ is not
an edge, otherwise $(x_i,x_{i+2})$ would be implied by the
 component $S_{x_ib}$. Since $S_{x_ib}$ is in $D \cup
N^+[S]$, the pair $(x_i,x_{i+2})$ is in $D \cup N^+[S]$, and we obtain a
circuit shorter than $C$. If $ax_{i+2}$ is an edge, then we have
$S_{ab} = S_{x_{i+2}b} = S_{x_{i+2}x_{i+1}}$, implying
$(x_{i+2},x_i)$ is in $D \cup N^+[S]$, a contradiction. So $ax_{i+2}$
is not an edge. Hence we have $S_{bg} = S_{x_{i+1}x_{i+2}} =
S_{ax_{i+2}} = S_{x_ix_{i+2}}$, a contradiction. Therefore, $x_i$
and $x_{i+2}$ have different colors.

Without loss of generality, we may assume that $x_i, x_{i+1}$ have the same
color for each even $i$.

Thus $(x_i,x_{i+1})$ is implied ( $N(x_{i}) \subseteq N(x_{i+1})$)
by a  component if and only if $i$ is even. We now proceed to
identify an exobiclique in $H$. Since the arguments are similar,
but there are many details, we organize the proof into small
steps. Note that by our assumption $x_{2i+1},x_{2i+2}$ have
different colors.

\begin{enumerate}
 \item Since $(x_{2i+1},x_{2i+2})$ is in a component $S_{2i+1}$ of $H^+$, by Lemma \ref{giantap}
there are two independent edges $x_{2i+1}a_i,x_{2i+2}b_i$ and $(a_i,b_i),(x_{2i+1},x_{2i+2}),(a_i,x_{2i+2})$ are in $S_{2i+1}$.

\item Since $(x_{2i},x_{2i+1})$ is implied by a  component
$S_{2i}$ od $H^+$, by Lemma \ref{implyap} there is an induced path
$x_{2i}, c_i, x_{2i+1}, e_i, d_i$ in $H$ satisfying the property
that $N(x_{2i}) \subseteq N(x_{2i+1})$ and $(c_i,e_i) \in S_{2i}$.

\item $a_ix_{2i+1},e_{i+1}d_{i+1}$ are independent edges of $H$.
This follows by applying a similar argument as in Lemma
\ref{transitiveap} for $S_{a_ix_{2i+2}},S_{x_{2i+2}d_{i+1}}$.
 Similarly $a_{i+1}x_{2i+3},d_{i+2}e_{i+2}$ are
independent edges of $H$

\item $x_{2i+2}x_j$, $ j \ne 2i+2$, is not an edge of $H$ as otherwise
$(x_{2i+1},x_{2i+2}) \rightarrow (x_{2i+1},x_j)$ and hence
$(x_{2i+1},x_j)$ is an implied pair by a component and we get a
shorter circuit using pair $(x_{2i+1},x_j)$.

\item $x_{2i+1}x_{2i+3}$ is an edge of $H$. Otherwise $x_{2i+1}a_i,x_{2i+3}b_i$ are independent edges and hence $(x_{2i+1},x_{2i+3})$ is in
the same component as $(x_{2i+1},x_{2i+2})$ and hence we get a shorter circuit by using pair $(x_{2i+1},x_{2i+3})$.

\item $x_{2i+1}b_{i+1}$ is an edge as otherwise $x_{2i+1}x_{2i+3},b_{i+1}x_{2i+4}$ are independent edges and hence $(x_{2i+3},x_{2i+4}),(x_{2i+1},x_{2i+4})$
are in the same component and we get a shorter circuit. Similarly $x_{2i+1}c_{i+2}$ is an edge of $H$.

\item $e_{i+2}x_{2i+1}$ is an edge as otherwise $(c_{i+2},e_{i+2}) \rightarrow (x_{2i+1},e_{i+2}) \rightarrow (x_{2i+1},x_{2i+5})$, a contradiction.
Unless $n=3$ and in this case by definition $e_{i+2}x_{2i+1}$ is an edge.

\item $b_ib_{i+1},c_{i+1}b_{i+1},c_{i+1}c_{i+2}$ are edges of $H$. These follow by applying Lemma \ref{new}
for $(x_{2i+1},b_i) \\ ,(x_{2i+3},b_{i+1})$,
and for $(x_{2i+1},b_{i}),(x_{2i+3},c_{i+1})$ and for
$(x_{2i+1},c_{i+1}),(x_{2i+3},b_{i+1})$ and for
$(x_{2i+1},c_{i+1}),\\ (x_{2i+3},c_{i+2})$.

\item $b_ie_{i+2}$, $c_{i+1}e_{i+2}$  are edges of $H$  because
$x_{2i+1}e_{i+2}$ is an edge of $H$ and hence we can apply Lemma \ref{new} for
$(c_{i+2},e_{i+2}),(x_{2i+1},b_i)$, and for
$(c_{i+_2},e_{i+2}),(x_{2i+1},c_{i+1})$. 

\item Analogous to (9) we conclude that $e_{i+2}e_{i+1}$ is an
edge of $H$.

\end{enumerate}

Now we have an exobiclique on the vertices
$$a_i,x_{2i+1},x_{i+2},b_i,c_{i+1},e_{i+1},d_{i+1},x_{2i+3},a_{i+1},b_{i+1},c_{i+2},x_{2i+4},d_{i+2},e_{i+2}.$$
Note that every vertex in $\{x_{2i+1},b_i,c_{i+1},e_{i+1}\}$ is
adjacent to every vertex in
$\{x_{2i+3},b_{i+1},c_{i+2},e_{i+2}\}$. Moreover by the assumption
and (3) $a_i,x_{2i+2},d_{i+1}$ have incomparable neighborhood in
$\{x_{2i+1},b_i,$ \\ $ c_{i+1},e_{i+1}\}$ and
$a_{i+1},x_{2i+4},d_{i+2}$ have incomparable neighborhood in
$\{x_{2i+3},b_{i+1},c_{i+2},e_{i+2}\}$.

\qed

Theorem \ref{trichotomyap} implies the correctness of  lines 4--12.
Specifically, we have the following Corollary.

\begin{cor}
If within lines 4--12 of the algorithm, we encounter a  component $S$
such that we cannot add either $N^+[S]$ or $N^+[S']$ to the current $D$,
then $H$ has an exobiclique.
\end{cor}
\pf
We cannot add $N^+[S]$ and $N^+[S']$ because the additions create
circuits in $D \cup N^+[S]$ respectively $D \cup N^+[S']$.
If either circuit leads to (iii) (in Theorem \ref{trichotomyap}) we are done by Theorem \ref{exohikoap}.
If both lead to (i) or (ii) (in Theorem \ref{trichotomyap}), we proceed as follows. Assume $(x_0,x_1), \dots, (x_n,x_0)$
is a shortest circuit created by adding $N^+[S]$ to the current $D$, and $(y_0,y_1),$
$\dots, (y_m,y_0)$ is a shortest circuit created by adding $N^+[S']$ to the current $D$.
We may assume that $N^+[S]$ contributes $(x_n,x_0)$ to the first circuit and $N^+[S']$
contributes $(y_m,y_0)$ to the second circuit. Note that $N^+[S]$ and $N^+[S']$ do not
contribute other pairs to these circuits. Indeed, if say
pairs $(x_n,x_0), (x_i,x_{i+1})$ are in the same component of $H^+$, then $x_n, x_i$
or $x_0, x_{i+1}$ are in the same $H_a$ by Remark \ref{pre-insect_compap}.

We assume each $x_i \in H_{a_i}$ and $y_j \in H_{b_j}$, thus all
pairs $(x_i,x_{i+1})$, $(y_j,y_{j+1})$ are in  components (not
implied by  components). Thus $S$ must contain both $(x_n,x_0)$
and $(y_0,y_m)$. If $Z$ is empty, we can conclude by Remark
\ref{pre-insect_compap} that $a_n=b_0$ and $a_0=b_m$, and
therefore $(x_{n-1},y_0)$, $(x_{n-1},x_n)$ are in the same
 component, and $(y_{m-1},y_m)$, $(y_{m-1},x_0)$ are
also in the same  component, and hence $(x_{n-1},y_0)$,
$(y_{m-1},x_0)$ are already in $D$. Therefore,
$$(x_0,x_1),(x_1,x_2), \dots, (x_{n-1},y_0),(y_0,y_1),\dots, (y_{m-2},y_{m-1}),(y_{m-1},x_0)$$
is a circuit in $D$, contrary to assumption. If $Z$ is non-empty, we can proceed in
exactly the same manner, knowing that no vertex $x_i$ or $y_j$ lies in $H_1$.
\qed

\section{Structure of a circuit after line 12 }\label{circuit-structure}

We consider what happens when a circuit is formed
during the execution of lines 13--21 of the algorithm. In what follows we specify the length and the properties of a circuit in $D$,
considering level by level construction of $N^*[D]$ (envelope of $D$). We break this section into two subsections. The first one proves that a minimal (we define 
what it means) circuit should have length at most four. In the second subsection we further analyze the pairs of each circuit and identify a dictator component associated 
to this circuit. 

\subsection { The length of a minimal circuit } 

\begin{dfn}
By a {\em minimal chain} between $x_0,x_n$ we mean the first time (the smallest level)
that there is a sequence $(x_0,x_1),(x_1,x_2),...,(x_{n-1},x_n)$ of the pairs in $D$ implying
$(x_0,x_n)$ in $D$ where none of the pairs $(x_i,x_{i+1})$, $0 \le i \le n-1$ is by transitivity.

Moreover, there is no $(x',y') \rightarrow (x_0,x_n)$ for
which the length of the minimal chain between $x'$, $y'$ is less than $n$.
\end{dfn}

\begin{dfn}
Let $C$ be a circuit in $N^*[D]$. We say $C$ is a {\em minimal circuit } if first, the latest pair in $C$ is created as early as possible (smallest possible level) 
 during the execution of $N^*[D]$;  second, $C$ has the minimum length; third, no pair in $C$ is by transitivity; finally, each pair is an original pair. 
\end{dfn}


\begin{lemma}\label{chainap}
Let $(x,y)$ be a pair in $D$ after line 12  of the algorithm, and current $D$ has no circuit.
Suppose $(x,y)$ is obtained by a minimal chain $CH : (x_0,x_1),(x_1,x_2),..., (x_{n-1},x_n),(x_n,x_{n+1})$ ($x_0=x$ and $x_{n+1}=y$). Then the following hold. 

\begin{enumerate}
\item $x_i,x_{i+2}$ have always different colors.
\item If $x,y$ have the same color then $n \le 3$ and $x_n,y$ have different colors.
\item If $x,y$ have different colors then $n \le 2$.
\begin{itemize}
\item If $n=2$ then $x_n,y$ have the same color.
\item If $n=1$ and $xy$ is not an edge then $x,x_1$ have the same color.
\item If $n=1$ and $xy$ is an edge then $x_1,y$ have the same color.
\end{itemize}
\end{enumerate}
\end{lemma}
{\textbf {Proof of 1.}} Suppose first all three $x_i, x_{i+1}, x_{i+2}$
have the same color, say black. 
Since $(x_i,x_{i+1})$ is not obtained by transitivity,
there exists a white vertex $a$ of $H$ such that
the pair $(x_i,a) \in D$ dominates $(x_i,x_{i+1})$ in $H^+$, i.e.
$a$ is adjacent in $H$ to $x_{i+1}$ but not to $x_i$. For a
similar reason, there exists a white vertex $b$ of $H$ adjacent to
$x_{i+1}$ but not to $x_i$, i.e. the pair $(x_{i+1},b) \in D$
dominates $(x_{i+1},x_{i+2})$ in $H^+$. 

We now argue that $a$ is not adjacent to $x_{i+2}$. Otherwise $(x_i,a) \in D$
also dominates the pair $(x_i,x_{i+2})$, and hence $(x_i,x_{i+2})$ is also in $D$ (at the same level as
$(x_i,x_{i+1})$), contradicting the minimality of $CH$.

Next we observe that the pair $(x_i,a)$ is not by transitivity.
Otherwise $(x_i,x_{i+1}), \\ (x_{i+1},x_{i+2})$ can be replaced by
a chain obtained from the pairs that implies $(x_i,a)$ together with
the pair $(a,x_{i+2})$. The pair $(a,x_{i+2})$ lies in the same
component of $H^+$ as $(x_i,x_{i+2}) \in D$ since the edges
$x_{i+1}a, x_{i+2}b$ are independent. Since all pairs of a
 component are chosen or not chosen for $D$ at the same
time, this contradicts the minimality of $CH$.  Thus, 
$(x_i,a)$ is dominated in $H^+$ by some pair $(c,a) \in D$. Since
$x_i, a$ have different colors, this means $c$ is a white vertex
adjacent to $x_i$. Note that $c$ is not adjacent to $x_{i+2}$,
as otherwise, $(c,a) \in D$ dominates $(x_{i+2},a)$, which would
place $(x_{i+2},a)$ in $D$, contrary to $(a,x_{i+2}) \in D$.

Now we claim that $b$ is not adjacent to $x_i$ in $H$. Otherwise the
pair $(x_{i+1},b) \in D$ dominates in $H^+$ the pair
$(x_{i+1},x_i)$, while $(x_i,x_{i+1}) \in D$, a circuit in $D$. Finally, $c$ is not
adjacent to $x_{i+1}$. Otherwise $cx_{i+1}, bx_{i+2}$ are
independent edges in $H$, and $cx_i, bx_{i+2}$ are also
independent edges in $H$; thus the pairs $(x_i,x_{i+2})$
and $(x_{i+1},x_{i+2})$ are in the same component, contradicting
again the minimality of $CH$. Now $(x_i,x_{i+1})$ and
$(x_{i+1},x_{i+2})$, $(x_i,x_{i+2})$  are in components. Since
there is no circuit in $D$, according to the rules of the
algorithm $(x_i,x_{i+2}) \in D$, contradicting the minimality of $CH$

We now consider the case when $x_i, x_{i+2}$ are black and $x_{i+1}$ is white.
As before, there must exist a white vertex $a$ and a black vertex $b$ such that
the pair $(a,x_{i+1})$ dominates $(x_i,x_{i+1})$ and the pair $(b,x_{i+2})$
dominates $(x_{i+1},x_{i+2})$; thus, $ax_i$ is an edge of $H$ and so is $bx_{i+1}$.
Note that the pair $(a,x_{i+1})$ dominates the pair $(x_i,x_{i+1})$ which dominates
the pair $(x_i,b)$. Therefore, we can replace $x_{i+1}$ by $b$ and obtain a chain
which is also minimal. Now $(b,x_{i+2})$ is by transitivity and we can replace it by a minimal chain. This would contradict the minimality of $CH$.

\begin{claim}
$ n \le 4$.
\end{claim}
{\textbf {Proof of the Claim.}}
Set $x_0=x$ and $x_{n+1}=y$. Let $i$ be the minimum number such
that $x_i, x_{i+1}$ have the same color, say black and
$x_{i+2},x_{i+3}$ are white. Let $x'$ be a vertex such that
$(x_i,x') \in D$ dominates $(x_i,x_{i+1})$. Note that if $x_{i+4}$
exists then it is black. If $x_{i+4}$ exists and $n \ge 5$ then
$x_{i+4}$ is white, and $x'x_{i+4}$ is not an edge as otherwise
$(x_i,x')\rightarrow (x_i,x_{i+4})$ and we get a shorter
chain. Now let $y'$ be a vertex such that $(x_{i+4},y') \in D$
dominates $(x_{i+4},x_{i+5})$. Now $y'x_{i+1}$ is not an edge as
otherwise $(x_{i+4},y')\rightarrow (x_{i+4},x_{i+1})$ and
we get a circuit
$(x_{i+1},x_{i+2}),(x_{i+2},x_{i+3}),(x_{i+3},x_{i+4}),(x_{i+4},x_{i+1})$ in $D$. 
Now $x'x_{i+1},y'x_{i+4}$ are independent edges and hence
$(x_{i+1},x_{i+4})$ is in a component. Note that each component
or its coupled is in $D$. $(x_{i+4},x_{i+1})$ is not in $D$ as
otherwise we get a circuit in $D$, and hence $(x_{i+1},x_{i+4})
\in D$, and we get a shorter chain. Thus we may assume that
$x_{i+4}$ does not exist. This means $x_{i+4}=y$. Now by
minimality assumption for $i$, $x_{i-1}= x_0$ and hence $n \le 4$. \\

{\textbf {Proof of 2.}} Suppose $x,y$ have the same color. We show that $n \le 3$. For contradiction suppose $n=4$. Now
according to (1) $x,x_1,x_4,y$ have the same color opposite to the
color of $x_2,x_3$. Let $y'$ be a vertex such that $(x_4,y')$ dominates $(x_4,y)$, and
let $x'$ be a vertex such that $(x_0,x') \in D$ dominates $(x_0,x_1)$.
Note that $y'x$ is not an edge as otherwise $(x_4,y') \rightarrow (x_4,x_0)$, implying a circuit in $D$. Similarly $x_1y$ is not an
edge of $H$. Finally $x'y$ not an edge as otherwise $(x,x') \rightarrow (x,y)$, contradiction to 
minimality of $CH$. Now $x_1x',y'y$ are independent edges and hence $(x_1,y)$ is in a component and hence
$(x_1,y) \in D$, contradicting the minimality of $CH$.
Therefore, $n \le 3$.

We continue by assuming $n=3$. We first show that
$x_3,y$ have different colors. In contrary suppose $x_3,y$ have
the same color. According to (1), $x_1,x_2$ have the same color
opposite to the color of $x,y,x_3$. Let $(x_1,x') \in D$ be a pair that dominates
$(x_1,x_2)$. Let $y''$ be a vertex such that $(x_3,y'')$
dominates $(x_3,y)$. $y''x$ is not an edge as otherwise
$(x_3,y'') \rightarrow (x_3,x)$ and we get a circuit. Let
$x''$ be a vertex such that $(x'',x_1) \in D$ dominates $(x,x_1)$.
Now $x' x''$ is not an edge as otherwise $(x_1,x')$ dominates
$(x'',x_1)$ and we get a circuit in $D$. We continue by having $x_2x$ as an edge of
$H$ as otherwise $x_2x',xx''$ are independent edges and hence
$(x,x_2)$ would be in a component that has already been placed in
$D$, contradicting the minimality of $CH$. Now
$(x_2,x_3),(x_3,y'')$ would imply $(x_2,y'')$ and $(x_2,y'')
\rightarrow (x,y'') \rightarrow (x,y)$. This would be a
contradiction to the minimality of $CH$. In fact we obtain
$(x,y)$ in less number of steps of transitivity. \\

{\textbf {Proof of 3.}} Suppose $x,y$ have different colors. We show that $n \le 3$. For contradiction suppose $n=4$.
Now according to (1) $x,x_3,x_4$ have the same color and opposite
to the color of $x_1,x_2,y$. We observe that
$xy$ is not an edge as otherwise $(x_4,y)$ would dominates
$(x_4,x)$ and hence we get a circuit in $D$. Let $x'$ be a vertex such that $(x_1,x') \in D$ dominates $(x_1,x_2)$ and $x''$ be a vertex
such that $(x'',x_1) \in D$ dominates $(x,x_1)$. Now $x'x''$ is
not an edge as otherwise $(x_1,x')$ dominates $(x'',x_1)$ and we
get a circuit in $D$. Moreover, $x_2x$ is an edge of $H$ as otherwise
$x_2x',xx''$ are independent edges and hence $(x,x_2)$ would be in
a component that has already been placed in $D$, contradicting
the minimality of $CH$. Now $(x_2,x_3),(x_3,x_4),(x_4,y)$
would imply $(x_2,y)$ and $(x_2,y)$ dominates $(x,y)$. This would
be a contradiction to the minimality of $CH$. In fact we
obtain $(x,y)$ in fewer number of transitivity applications.

Therefore, $n \le 3$. Now it is not difficult to see that either
$n=2$ and $x,x_1$ have the same color opposite to the color of
$x_2,y$ or $n=1$.

Suppose $n=1$.  
First assume $xy$ is an edge. Now $x_1,y$ have the same color as otherwise $(x_1,y) \rightarrow (x_1,x)$, a contradiction.

Thus we continue by assuming $xy$ is not an edge. We show that $x_1,x$ have the same color. 
For contradiction suppose $x_1,y$ have the same color. 
Let $(x',x) \in D$ be a pair that dominates $(x,x_1)$ and let $(x_1,y') \in D$ be a pair that dominates $(x_1,y)$. Now $x'y'$ is
not an edge and hence, $yy',xx'$ are independent edges. This
shows that $(x,y)$ is in a component, contradicting the
minimality of $CH$. \qed

\begin{cor}\label{chain1ap}
Let $(x,y)$ be a pair in $D$ after line 12 of the algorithm, and current $D$ has no circuit.

\begin{itemize}
\item Suppose $x,y$ have the same color and $(x,w) \rightarrow (x,y)$ such that $(x,w)$ is by transitivity with a minimal 
chain $(x,w_1),(w_1,w_2),\dots,(w_m,w)$.
Then $m=2$ and $x,w_1$ have the same color and opposite to the color of $w_2,w$.

\item Suppose $x,y$ have different colors and $(w,y) \rightarrow (x,y)$ such that $(w,y)$ is in a trivial component. Then
$(w,y)$ is by transitivity with a minimal chain
$(w,w_1),(w_1,w_2),(w_{2},y)$ where $w_1,w_2$ have the same color
opposite to the color of $w,y$.

\end{itemize}
\end{cor}
\pf If $x,y$ have the same color then by Lemma \ref{chainap} we have $m=2$ or $m=1$.
If $m=2$ then $x,x_1$ have the same color and opposite to the color of $x_2,w$. When $m=1$ then by Lemma \ref{chainap} (3), $w_1,y$ have the same color. 
Note that $(w_1,w)$ dominates $(w_1,y)$ and $(w_1,y)$ is in $N^*[D]$ at
the same time $(w_1,w)$ placed in $D$. Therefore, we use the chain $(x,w_1),(w_1,y)$ in order to obtain $(x,y)$, contradiction. 
If $x,y$ have different colors then by Lemma \ref{chainap} either $m=2$ or $m=3$.
If $m=3$ then $w,w_1,y$ have the same color and opposite to the color of $w_2,w_3$. Let $w'$ be a vertex such that $(w,w') \in D$ dominates $(w,w_1)$. 
We observe that $w_1,x$ is not an edge as otherwise $(w_1,y) \rightarrow (x,y)$
and hence, we obtain $(x,y)$ in an earlier level or in fewer steps of transitivity 
application because $(w_1,w_2),(w_2,w_3),(w_3,y)$ are in $N^*[D]$. Now $wx,w_1w'$ are independent edges and hence $(x,w_1)$ is already in $D$, 
so we may use the chain $C=(x,w_1),(w_1,w_2),(w_2,w_3),(w_3,y)$. Now by considering the chain 
$C$ we would obtain $(x,y)$ in some earlier step since $w_1,w_2$ have different colors, and this is a contradiction 
by Lemma \ref{chainap} (1). Therefore, $n=2$ and Lemma \ref{chainap} is applied.  \qed \\

Now by Lemma \ref{chainap} and Corollary \ref{chain1ap} we have the following.

\begin{cor}\label{circuit-lengthap}
Let $C=(x_0,x_1),(x_1,x_2),...,(x_{n-1},x_n),(x_n,x_0)$ be a minimal circuit, formed at lines 13--21 of the Algorithm. 
Then $n=3$ and $x_0,x_3$ have the same color and opposite to the color of $x_1,x_2$.
\end{cor}
\pf We may assume that non of the pair $(x_i,x_{i+1})$ in $C$ is by 
transitivity as otherwise we replace $(x_i,x_{i+1})$ by a minimal chain between $x_i,x_{i+1}$. 
Now we just need to apply Lemma \ref{chainap} and Corollary \ref{chain1ap}.
\qed
\vspace{1cm}

Therefore, in what follows, we may assume a minimal circuit $C$ has
the following form.

$$ C=(x_0,x_1),(x_1,x_2),(x_2,x_3),(x_3,x_0) \  \ , \  \  x_0,x_3  \  \  are \  \  white ,  \  \  x_1,x_2  \  \  are  \  \  black  \  \ vertices. $$

\subsection{Decomposition of the pairs of a minimal circuit and finding dictator components }

\begin{lemma}\label{cl1ap}
If $(x_1,x_2)$ ( $(x_3,x_0)$ ) is not simple then $(x_1,w)$ where $(x_1,w) \rightarrow (x_1,x_2)$ ( $(u,x_3)$ where $(u,x_3) \rightarrow (x_2,x_3)$ ) is by transitivity.
\end{lemma}
\pf For contrary suppose $(x_1,w)$ is not by transitivity. Thus
there is some $(w',w) \in N^*[D]$ such that $(w',w) \rightarrow (x_1,w)$. 
Now $(w',w)$ is not in a component as otherwise $(x_1,x_2)$ is implied by $S_{w'w}$
and hence $(x_1,x_2)$ is simple. Thus $(w',w)$ is by transitivity
and by Corollary  \ref{chain1ap} there are white vertices $w'_1,w'_2$ such that
$(w',w'_1),(w'_1,w'_2),(w'_2,w)$ are placed in $N^*[D]$ and they imply
$(w',w)$. Now $(x_0,x_1)$ and $(x_0,w')$ are placed in $N^*[D]$ at the same
time ($(x_0,x_1)\rightarrow (x_0,w')$). Moreover $(w'_2,w) \rightarrow
(w'_2,x_2)$, and they are placed in $N^*[D]$ at the same time(level). Now we would
have the circuit $(x_0,w'),(w',w'_1),(w'_1,w'_2),(w'_2,x_2), (x_2,x_3), (x_3,x_0)$, in $N^*[D]$ that 
contradicts the minimality of circuit $C$.\qed \\

In the rest of the proof we often use similar argument as in the
Lemma \ref{cl1ap} and we do not repeat it again. \\

{\textbf {Decomposition of each pair $(x_i,x_{i+1}) \in C$ and associating a component $S_i$ to $(x_i,x_{i+1})$ }}\\

In what follows we decompose each of the pairs $(x_0,x_1),(x_1,x_2),(x_2,x_3),(x_3,x_0)$, meaning that we analyze the steps in
computing $N^*[D]$ to see how we get these pairs. If $(x_i,x_{i+1})$ is simple then there exists a component $S_i$ such 
that $(x_i,x_{i+1}) \in N^+[S_i]$ and hence $DCT(x_i,x_{i+1})=S_i$. 

When $(x_i,x_{i+1})$ is a complex pair and $x_i,x_{i+1}$ have
the same color then by Lemma \ref{cl1ap} there is a pair  $(x_i,w) \in N^*[D]$ such
that  $(x_i,w) \rightarrow (x_i,x_{i+1})$ and $(x_i,w)$ is by
transitivity over $(x_i,w_1),(w_1,w_2),\\ (w_2,x_{i+1}) \in N^*[D]$ where
$w_2,w$ have the same color and opposite to color of
$x_i,x_{i+1},w_1$.  Now if $(x_i,w_1) \in N^+[S]$ for some component
$S$ then we set $S_i=S$ otherwise we recursively decompose $(x_i,w_1)$ in
order to obtain $S_i$. 

The goal is to show that when
$(x_i,x_{i+1})$ and $(x_j,x_{j+1})$ are both complex then
$S_i=S_j$ and $S_i$ is the dictator component. Moreover, we show
that if $(x_1,x_2)$ is a complex pair and $(x_0,x_1)$ is in a
 component then $(x_0,x_1) \in S_1$ and $S_1$ is the
dictator component. Similarly if $(x_3,x_0)$ is a complex pair and
$(x_2,x_3)$ is in a  component then $(x_2,x_3) \in S_3$ and $S_3$
is the dictator component.

If $(x_i,x_{i+1})$ is a simple pair in $N^+[S_i]$  and
$(x_j,x_{j+1})$ is a complex pair for some $0 \le j \le 3$ such
that $S_i \ne S_j$ then we show that by replacing $S_i$ with $S'_i$
 and keeping $S_j$ in $D$ we still get a
circuit in the envelope of $D$. \\

\textsl{ Before we continue we observe that $x_0x_2$ is an edge of $H$}.

For contrary, suppose $x_0x_2$ is not an edge of $H$.  Let $(p,x_1)$ be a pair
in $D$ that dominates $(x_0,x_1)$ ($(x_0,x_1)$ is not by
transitivity). Now $wp$ is not an edge as otherwise $(x_1,w)$
would dominates $(x_1,p)$ implying an earlier circuit in $D$. Now
$px_0,wx_2$ are independent edges and hence $(x_0,x_2)$ would be
in a  component and consequently $(x_0,x_2)$ has been already
placed in $D$ (if $(x_2,x_0)$ is in $D$ then we would have an
earlier circuit) implying a shorter circuit. Therefore, $x_0x_2$ is
an edge. \\

{\textbf {Assumption : }} Let $(x_1,w) \in N^*[D]$ such that $(x_1,w) \rightarrow (x_1,x_2)$ and $(u,x_3) \in N^*[D]$ such that $(u,x_3) \rightarrow (x_2,x_3)$ 
and $(x_3,v) \in N^*[D]$  such that $(x_3,v) \rightarrow (x_3,x_0)$.
In what follows we consider the decomposition of each of the pairs $(x_1,w)$ , $(u,x_3)$, and $(x_3,v)$. Keeping in mind that they are the earliest
pairs added into $N^*[D]$. \\

\hspace{5cm} {\textbf {Decomposition of $(x_1,w)$ }} \\

Suppose $(x_1,x_2)$ is not a simple pair. Then by Lemma \ref{cl1ap}
$(x_1,w)$ is by transitivity and hence by Lemma \ref{chainap}
there are vertices $w^1_1,w^1_2,w^1$, $w^1=w$ such that the chain
$(x_1,w^1_1),(w^1_1,w^1_2),(w^1_2,w^1) \in N^*[D]$ is minimal and imply
$(x_1,w^1)$. Moreover $w^1_1,w^1$ are white and $x_1,w^1_1$ are
black, and none of the $w^1_1w^1_2, x_1w^1$ is an edge of $H$. In
general suppose $(x_1,w^1)$ is obtained after $m$ steps (the minimum possible steps); meaning
that $(x_1,w^1_1)$ is not simple and is obtained after $m-1$ steps
of implications ( transitivity and out-section). \\

\begin{figure}[htbp] 
 \begin{center}
 \includegraphics[scale=0.7]{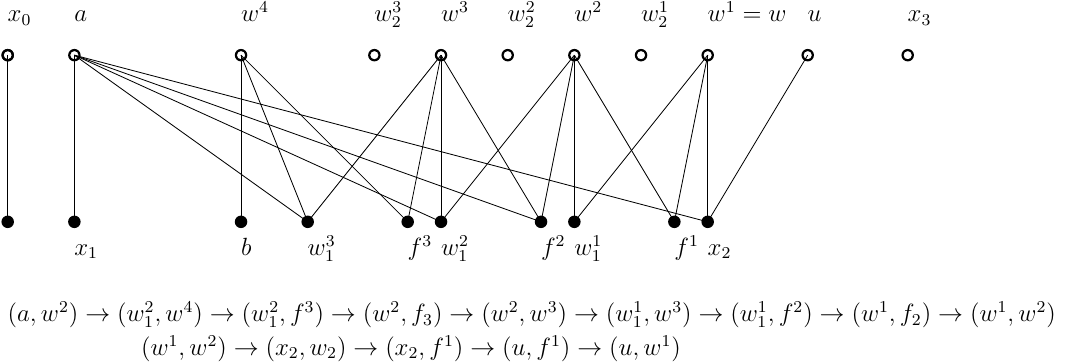}
 \caption{An example for decomposition of  $(x_1,w)$ }
 \label{fig:fig4}
 \end{center}
 \end{figure}

To summarize:  for every $1 \le i \le m-1$ we have the following (see Figure \ref{fig:fig4}). 

\begin{enumerate}
\item[1.] By similar argument as in Lemma \ref{cl1ap}, $(x_1,w^i)$ is by
transitivity and hence there are vertices $w^i,w^i_1,w^i_2$ such
that $(x_1,w^i_1),(w^i_1,w^i_2),(w^i_2,w^i) \in N^*[D]$ is a minimal chain
and implies $(x_1,w^i)$.

\item [2.] $w^i,w^i_2$ are white and $w^i_1$ is
black.

\item [3.]  $(x_1,w^{i+1}) \in N^*[D]$ and $(x_1,w^{i+1}) \rightarrow (x_1,w^i_1)$.

\item [4.] None of the $w^i_1w^i_2, x_1w^i$ is an edge of $H$.

\end{enumerate}

Since $m$ is minimum,  we have the following :

\begin{enumerate}

\item [5.] $w^i_2x_2$ is not an edge of $H$, as otherwise $(w^i_1,w^i_2) \in N^*[D]$,  $(w^i_1,w^i_2) \rightarrow (w^i_1,x_2)$ and 
hence, we get $(x_1,x_2)$ earlier because of the earlier chain $(x_1,w^i_1),(w^i_1,x_2) \in N^*[D]$, a contradiction. 

\item [6.] There is no edge from $w^{i+1}$ to $w^{i-1}_1$. Otherwise $(x_1,w^{i+1})\rightarrow(x_1,w^{i-1}_1)$ and
hence, we get a shorter chain $(x_1,w^{i-1}_1),(w^{i-1}_1,w^{i-1}_2),(w^{i-1}_2,w^{i-1}) \in N^*[D]$, and
consequently we obtain $(x_1,w^1)$ in less than $m$ steps. 

\item [7.] There are vertices $f^i$, $1 \le i \le m-1$  such that $(w^i_2,f^i) \rightarrow (w^i_2,w^i)$.

\item [8.] $f^iw^j$, $ j \le i+1$ is not an edge of $H$ as otherwise $(w^{i}_2,f^i)\rightarrow (w^i_2,w^j)$ and hence, we
use the chain $(x_1,w^i_2),(w^i_2,w^j)$ to obtain $(x,w^1)$ in less than $m$ steps. Similarly
$f^iw^j_2$, $j \le i+1$ is not an edge. 

\item [9.] $w^i_2x_1$ is not an edge as otherwise $(w^i_1,w^i_2) \rightarrow (w^i_1,x_1)$ 
and hence, we get an earlier circuit because $(x_1,w^i_1)$ is in $N^*[D]$.

\item [10.] $w^i_1w^i$ is an edge as otherwise
$w^i_1w^{i+1},w^iw^{i-1}_1$ are independent edges and hence
$(w^{i+1},w^{i-1}_1)$ is in a component already placed in $D$
(otherwise we would have an earlier circuit using
$(x_1,w^{i+1}),(w^{i+1},w^{i-1}_1),(w^{i-1}_1,w^{i-1}_2)....))$, a
contradiction to the minimality of the chain implying $(x_1,w^1)$.

\item [11.] There are vertices $a,b$ such that $x_1a$ and $w^mb$ are
independent edges; $(x_1,w^m)$ is in a component $S_1$. This is because $(x_1,w^m)$ is simple and $x_1,w^m$ have different colors.

\item [12.] $w^mw^j_1$ is not an edge of $H$  for $j < m$.   Otherwise $(x_1,w^m) \rightarrow (x_1,w^j_1) \in N^*[D]$ and hence we get a shorter chain
$(x_1,w^j_1),(w^j_1,w^j_2),(w^j_2,w^j)$, and we get $(x_1,w)$ in less than $m$ steps.

\end{enumerate}

When $m \ge 3$ we have the following. 

\begin{enumerate}
\item [13.] $aw^{m-2}_1$ and $af^{m-2}$ are edges of $H$. Otherwise because $w^{m-1}x_1$ is not an edge, $(x_1,w^{m-2})$ is in a component that
 placed in $D$ ( since we are after line 12 , and
$(w^{m-2},x_1)$ is not in $D$ as otherwise it would yield an earlier
circuit) implying $(x_1,w)$ in less than $m$ steps.

\item [14.] $f^{i}w^{i+1}$ is an edge of $H$. Otherwise $(w^i_1,w^i_2),(w^i_2,f^{i}) \in N^*[D]$ imply $(w^i_1,f^{i}) \in N^*[D]$ and
now $(w^i_1,f^{i}) \rightarrow (w^{i+1},f^{i}) \rightarrow (w^{i+1},w^i)$ and hence $(w^{i+1},w^i) \in N^*[D]$.
This would contradict the minimality of the chain, fewer number of steps in obtaining $(x_1,w^1)$. 

\item [15.] We have $(a,w^m) \rightarrow (w^{m-2}_1,w^m) \rightarrow
(w^{m-2}_1,f^{m-1})$ and $(w^{m-2}_1,f^{m-1}) \rightarrow (w^{m-2},w^{m-1})$. Moreover, we see that $(w^{i-1},w^i)
\rightarrow (w^{i-2}_1,f^{i-1}) \rightarrow (w^{i-2},w^{i-1})$, $3
\le i \le m-1$.

\item [16.] $(w^1,w^2) \rightarrow (x_2,w_2) \rightarrow (x_2,f_1)$. This would imply that there exists a path from both $(x_1,b),(x_1,w^m)$ to $(x_2,f^1)$. 

\end{enumerate} 


%

\begin{figure}[htbp] 
 \begin{center}
 \includegraphics[scale=0.5]{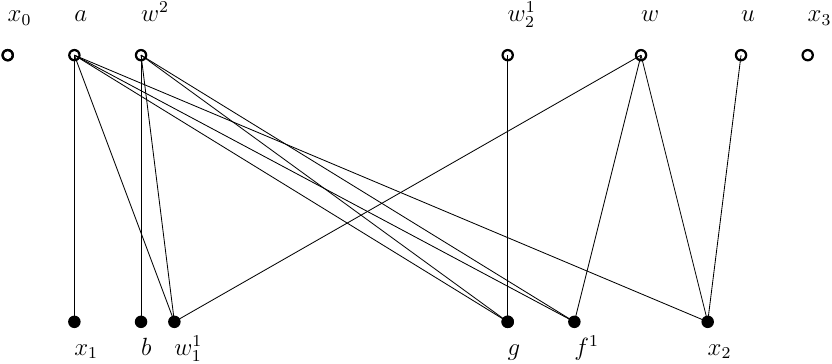}
 \caption{An example for decomposition of  $(x_1,w)$ }
 \label{fig:fig5}
 \end{center}
 \end{figure}

Now we suppose $m=2$. In this case by similar line of reasoning as above we have the following (see Figure \ref{fig:fig5}). 

\begin{enumerate} 

\item [17.] Observe that $bw^1 \not\in E(H)$. Otherwise $(x,b) \rightarrow (a,b) \rightarrow (a,w^1) \rightarrow (x_1,w^1)$. This is a contradiction to 
the minimality of the chain implying $(x_1,w)$. 

\item [18.] Let $g^1$ be a vertex in $H$ such that $(g^1,f^1) \in N^*[D]$ and $(g^1,f^1) \rightarrow (w^1_2,f)$. 
We have $w^2g^1 \in E(H)$. Otherwise since $g^1w^1_2,w^2w^1_1$ are edges of $H$ 
and $w^1_1w^1_2 \not\in E(H)$, $g^1w^1_2,w^2w^1_1$ are independent edges, and hence, 
we obtain $(x_1,w^1)$ by chain $(x_1,w^2),(w^2,w^1_2),(w^1_2,w^1)$, a contradiction (to minimality of $(x_1,w^1)$).

\item [19.] $w^2f^1$ is an edge of $H$. Otherwise $(g^1,f^1) \rightarrow (w^2,f^1) \rightarrow (w^2,w) \rightarrow (b,w)$. Therefore $S_{bw} \subset D$ and hence 
we can use the chain $(x_1,b),(b,w^1)$ to obtain $(x_1,w^1)$, a contradiction.

\item [20.]  $ax_2$ is not an edge of $H$. Otherwise $x_1a,x_2w^1$ are independent edges and hence $(x_1,x_2)$ is a simple pair ($S_{x_1x_2}$ is a component). 

\item [21.]  $w^2x_2$ is not an edge. Otherwise $(x_1,w^2) \rightarrow  (x_1,x_2) \in N^*[D]$ a contradiction to the minimality of chain $(x,w^1)$.

\item [22.]  $(a,w^2) \rightarrow (x_2,w^2) \rightarrow (x_2,f^1) \in N^*[D]$.


\end{enumerate}

{\textbf {Conclusion}}  We have the following. 
\begin{itemize}
  \item $f^1u$ is not an edge of $H$. Otherwise $(w^1_2,f^1)\rightarrow(w^1_2,u) \in N^*[D]$ and we get an earlier
circuit $(x_1,w^1_1),(w^1_1,w^1_2),(w^1_2,u),(u,x_3)$.
\item $(x_2,f^1) \rightarrow (u,f^1) \rightarrow (u,w^1)$ because $f^1u$ is not an edge

\item $(x_2,f^1),(x_1,w^m) \in S_1$ 
 \end{itemize}

\hspace{5cm} {\textbf {Decomposition of $(u,x_3)$ }}\\

\begin{figure}[htbp] 
 \begin{center}
 \includegraphics[scale=0.65]{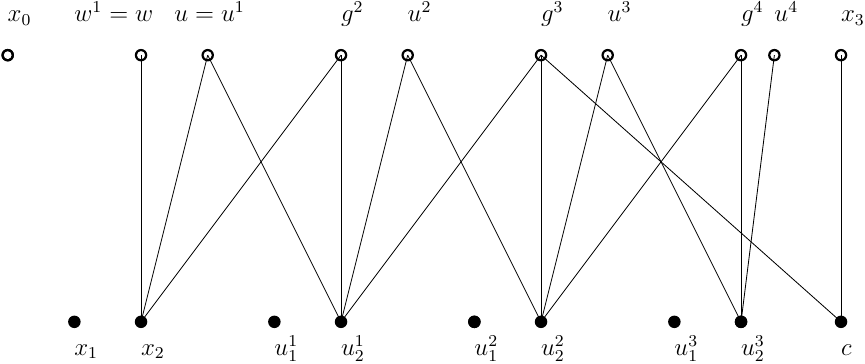}
 \caption{An example for decomposition of  $(u,x_3)$ }
 \label{fig:fig7}
 \end{center}
 \end{figure}

Suppose $(x_2,x_3)$ is a complex pair. Then $(u,x_3)$ is obtained after $n >1 $ steps as follows (see Figure \ref{fig:fig7}). 

There are vertices $u^1=u$ and $g^i,u^i,u^i_1,u^i_2$, $1 \le i \le n$,  such that :

\begin{enumerate}
\item[1.] $(u^i,x_3) \rightarrow (u^{i-1}_2,x_3)$, $i \ge 2$. 
\item[2.] $(u^i,u^i_1),(u^i_1,u^i_2),(u^i_2,x_3)$ imply $(u^i,x_3)$. Moreover, $u^i$
is white and $u^i_1,u^i_2$ are black. 
\item[3.] for $1 < i \le n$, $u^iu^{i-1}_2$ is an edge. 
\item[4.] $(u^{n-1}_2,x_3)$ is in a component. 
\item[5.] $(u^i_1,g^{i+1}) \in N^*[D]$ where $(u^i_1,g^{i+1}) \rightarrow (u^i_1,u^i_2)$ for $1 < i \le n-1$. 
\item[6.] There is a vertex $c$
such that $u^nu^{n-1}_2,cx_3$ are independent edges of $H$, and
$(u^{n-1}_2,x_3), (u^n,x_3)$ are in a component $S_2$. This follows because $(u^{n-1}_2,x_3)$ is simple and $u^{n-1}_2,x_3$ have different colors.

\end{enumerate}

\begin{enumerate}

\item [7.] $(u^{n-1}_2,x_3), (u^n,x_3)$ are in
the component $S_2$.

\item [8.] $u^i,u^{j}_2$, $j \le i-2$ are not adjacent. Otherwise $(u^i,x_3) \rightarrow (u^{j}_2,x_3)$ and
hence we get $(u,x_3)$ in less than $n$ steps.

\item [9.] $u^iu^i_2$ is
an edge of $H$. Otherwise $u^iu^{i-1}_1, u^{i+1}u^i_2$ are independent
edges and hence $(u^{i-1},u^{i})$ is in a  component placed in
$D$, implying an earlier (shorter) chain.

\item [10.] By definition
$(u^i_1,g^{i+1}) \rightarrow (u^i_1,u^i_2)$ 


\item [11.] $u^{i+1}u^{i-1}_2$ is not an edge of $H$. Otherwise
$(u^{i+1},x_3) \rightarrow (u^{i-1}_2,x_3)$ and hence we would
obtain $(u,x_3)$ in less than $n$ steps.

\item [12.] $g^iu^i_2$ is not an edge of $H$. Otherwise $(u^{i-1}_1,g^{i}) \rightarrow (u^{i-1}_1,u^i_2)$ and
hence we obtain $(u,x_3)$ in less than $n$ steps. 

\item [13.] $x_2u^2$ is not an edge of $H$. This follows by similar argument as in (12). 

\item [14.] $u^{i-1}_2g^{i+1}$ is an edge of $H$. Otherwise
$(u^i,u^i_1),(u^i_1,g^{i+1}) \in N^*[D]$ would imply $(u^i,g^{i+1}) \in N^*[D]$ and as
consequence $(u^i,g^{i+1}) \rightarrow (u^{i-1}_2,g^{i+1})
\rightarrow (u^{i-1}_2,u^i_2) \in N^*[D]$ and therefore we obtain
$(u,x_3)$ in less than $n$ steps. 

 \item [15.]  $x_2g^2$ is an edge. Otherwise we use the chain $(u^1,u^1_1),(u^1_1,g^1) \in N^*[D]$ imply $(u^1,g^1) \in N^*[D]$
 and hence $(u^1,g^1) \rightarrow (x_2,g^1) \rightarrow (x_2,u^1_1)$. Now $x_2w^1,u^1_1g^1$ are independent edges and hence $S_{x_2 u^1_1}$ 
is a component in $D$. This would contradict the minimality of chain obtaining $(x_2,x_3)$.

\item [16.] For every $2 \le i \le n-2$, we have $(u^i,g^i) \rightarrow
(u^i_2,g^i) \rightarrow (u^i_2,u^{i-1}_1) \rightarrow
(u^{i+1},u^{i-1}_1) \rightarrow (u^{i+1},g^{i+1})$.

\item [17.]  Observe that $w^1u^i_2$ is not an edge of $H$.  Otherwise $(x_1,w^1) \rightarrow
(x_1,u^i_2)$ and this contradicts the minimality of circuit $C$.

\item [18.] We have $(u^1,w^1) \rightarrow (u^1_2,w^1) \rightarrow
(u^1_2,x_2) \rightarrow (u^2,x_2) \rightarrow (u^2,g^2)$. 

\item [19.]  $g^{n-1}c$ is an edge of $H$. Otherwise $u^{n-2}_2g^{n-1},cx_3$
are independent edges an hence $(u^{n-2}_2,x_3)$ is in a component
and $(u,x_3)$ is obtained in less than $n$ steps.
\end{enumerate}



%

{\textbf{Conclusion :}} 
$(u^{n-1},g^{n-1}) \rightarrow (u^{n-1}_2,g^{n-1}) \rightarrow (u^{n-1}_2,c)$ \\

\hspace{5cm} {\textbf {Decomposition of $(x_3,v)$ }} \\

Suppose $(x_3,x_0)$ is a complex pair and it is obtained after $t$ steps. This means there are vertices $v^i,v^i_1,v^i_2$ for $1 \le i \le t$ and $v^1=v$ such that :

\begin{enumerate}
\item $(x_3,v^{i+1})$ implies $(x_3,v^i_1)$ \item
$(x_3,v^i_1),(v^i_1,v^i_2),(v^i_2,v^i)$ imply $(x_3,v^i)$.
$v^i,v^i_2$ are black and $v^i_1$ is white. \item $v^iv^{i-1}_2$,
$ 2 \le i \le t$ is an edge. \item $(x_3,v^t)$ is in a  component,
and $v^t$ is black.
\end{enumerate}

There are vertices $d,e$ such that $x_3d,v^te$ are independent edges and $v^tv^{t-1}_1$ is an edge. Let $S_3=S_{ex_3}$. Note that
$dv^{t-1}_1$ is also an edge. Let $g^{t-1}$ be a vertex that $(v^{t-1}_2,g^{t-1})$ implies $(v^{t-1}_2,v^{t-2})$.
As we argued in the decomposition of $(x_1,w^1)$,  $g^{t-1}v^t$ is an edge of $H$. We note that $dg^{t-1}$ is an edge as
otherwise since $x_3v^{t-1}$ is not an edge, $x_3d,v^{t-1}g^{t-1}$ are independent
edges and we obtain $(x_3,v)$ in less than $t$ steps. We also note that $v^tx_0$ is not an edge of $H$.

\begin{lemma}\label{relationship}
After considering the decomposition of each pairs of circuit $C=(x_0,x_1),(x_1,x_2),(x_2,x_3),(x_3,x_0)$ we have the following. 
\begin{enumerate}

\item If $(x_1,x_2)$
is a complex pair and $(x_2,x_3)$ is also a complex pair then
$S_1=S_2$. 

\item If $(x_1,x_2)$ is a complex pair and $(x_0,x_1)$ is in a
component $S_0$ then $(x_0,x_1) \in S_1$ and hence $S_0=S_1$ 

\item If $(x_2,x_3)$ is
a complex pair and $(x_3,x_0)$ is a simple pair implied by
 component $S_3$ then $S_3=S_2$ ($S_3$ is associated with pair $(x_3,x_0)$). 
 
\item If $(x_2,x_3)$ and $(x_3,x_0)$ are complex pairs
then $S_2=S_3$. 

\item If $(x_1,x_2)$ and $(x_3,x_0)$ are complex
pairs and $(x_0,x_1)$, \\ $(x_2,x_3)$ are simple pairs then
$S_1=S_3$ and $(x_2,x_3),(x_0,x_1) \in S_1$.
\end{enumerate}
\end{lemma}
\noindent

\begin{figure}[htbp] 
 \begin{center}
 \includegraphics[scale=0.65]{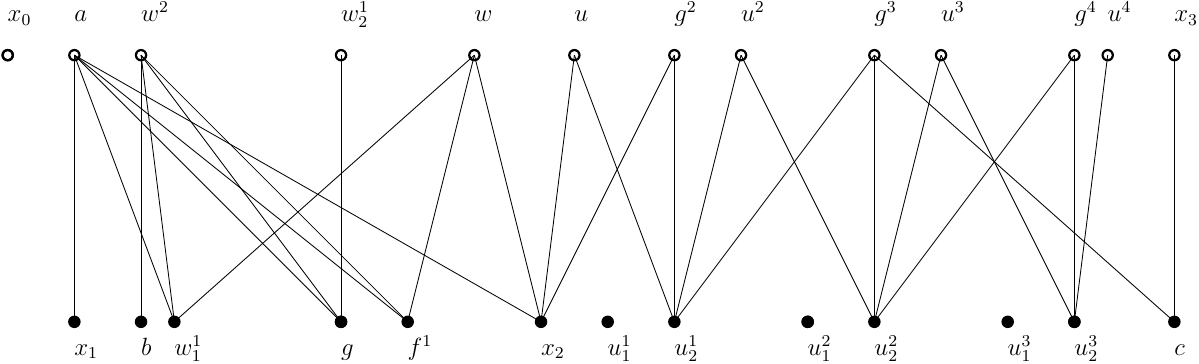}
 \caption{Illustration of the proof of $S_1=S_2$}
 \label{fig:fig8}
 \end{center}
 \end{figure}

\noindent
\textbf{Proof of 1:} We need to observe that there is a directed path from $(x_1,w^m) \in S_1$ to $(u^1,w^1)$ in $H^+$ 
Moreover, there is a directed path from $(u^1,w^1)$ to $(u^{1}_1,x_2)$ in $H^+$ (see the conclusion at the end of 
decomposition of $(x_1,w^1)$). 
There is also a direct path from $(u^1_2,x_2)$ to $(u^n,x_3)$ in $H^+$ (see the conclusion at the end of 
decomposition of $(u^1,x_3)$). 
We need to observe that $(u^1_2,x_2) \in S_1$ and $(u^n,x_3),(u^{n-1}_2,c) \in S_2$ and since there is a direct path 
from $S_1$ to $S_2$, $S_1=S_2$ (see Figure \ref{fig:fig8}).\\

\noindent
\textbf{Proof of 2:} Since $x_0,x_1$ have different colors, there are vertices $p,q$ such that $x_0p,x_1q$ are independent edges.
 Observe that $x_0f^{m-1}$ is not an edge as otherwise $(w^{m-1}_2,f^{m-1}) \rightarrow (w^{m-1}_2,x_0)$ and hence we get an earlier circuit
 $(x_0,x_1),(x_1,w^{m-1}_1) \\ ,(w^{m-1}_1,w^{m-1}_2),(w^{m-1}_2,x_0)$.

 Note that $pw^1$ is not an edge as otherwise $(p,x_1)$ dominates $(w^1,x_1)$ while $(x_1,w^1)$ is in $D$. Recall that $x_0x_2$ is an edge of $H$.
 Observe that $qx_2,ax_2$ are edges of $H$ as otherwise $(x_1,x_2)$ would be in a
 component, and is not complex.

 Also $qf^{m-1},af^{m-1}$ are both edges of $H$ as otherwise $x_1q,f^{m-1}w^{m-1}$ are independent edges and hence $(x_1,w^{m-1})$ is in a  component, and we obtain $(x_1,x_2)$ in less than $m$ steps.

Recall that $(x_0,x_1),(x_1,b) \in S_0$ and $(p,q),(x_1,w^m) \in
S_2$. If both $ap,qb$ are edges of $H$ we have $(x_1,b)
\rightarrow (a,b) \rightarrow (a,q) \rightarrow (p,q) \in S_0$ and
hence  $S_0=S_1$ (by the comment after the Corollary \ref{source-sinkap}) and claim is proved.

Therefore, we may assume at least one of  $qb,ap$ is not an edge
of $H$. We prove the claim for $qb$ not being an edge of $H$ and
the proof for $ap \not\in E(H)$ is similar. When $qb$ is not an
edge of $H$ $qx_1,bw^m$ are independent edges and hence $(q,w^m)
\in S_2$. Now we need to see that $x_0x_2,qx_2$ are edges of $H$
while $w^mx_2$ is not an edge of $H$ and
 $f^{m-1}w^m,f^{m-1}q $ are edges of $H$ while $x_0f^{m-1}$ is not an edge. These would imply that $(q,w^m) \rightarrow (x_2,w^m), \rightarrow (x_2,f^{m-1}) \rightarrow (x_0,f^{m-1}) \rightarrow (x_0,q) \in
 S_0$, and hence $(x_0,x_1), (x_1,w^m),(q,w^m)$ are in the same component $S_0=S_1$. \\

\noindent
\textbf{Proof of 3:} Since $(x_3,x_0)$ is implied by $(x_3,v)$ and $(x_3,x_0)$ is
simple, $(x_3,v)$ is in a  component and there are independent
edges $x_3c,vd$ of $H$.
  Note that $u^{n-1}v$ is not an edge as otherwise $(x_3,v) \rightarrow (x_3,u^{n-2})$ while $(u^{n-2},x_3) \in D$. 
  However $cu^{n-1}$ is an edge as otherwise $u^{n-1} u^{n-2}_2,x_3c$ are independent edges 
  and hence $(x^{n-2}_2,x_3)$ is in $D$, a contradiction.
  Now $u^{n-1}_2u^{n-1},cu^{n-1}$ are edges of $H$. We note that $w^1c$ is an 
  edge as otherwise $x_2w^1,x_3c$ are independent edges and hence $(x_2,x_3)$ 
  would be in a component, a contradiction to the assumption in (2). We show that $w^1v$ is an edge as otherwise
 $(d,v) \rightarrow (w^1,v) \rightarrow (w^1,x_0)$ and hence $(w^1,x_0) \in D$ 
 while $(x_0,x_1),(x_1,w^1)$ are also in $D$, yielding an earlier circuit in $D$.
Recall that $w^1u^{n-1}_2$ is not an edge. Now $cu^{n-1},u^{n-1}_2u^{n-1}$ are edges of $H$ 
while $vu^{n-1}$ is not an edge and $w^1c,w^1v$ are edges of $H$ while $w^1u^{n-1}_2$ is not an edge. 
These imply that $(x_3,v)$ and $(u^n,x_3)$ are in the same component $S_2$.  \\


\noindent \textbf{Proof of 4:} Observe that $g^{t-1}u^{n-1}_2$ is
not an edge as otherwise $(v^{t-1}_2,g^{t-1}) \rightarrow 
(v^{t-1}_2,u^{n-1}_2)$ and now we have an earlier circuit
$(u^{n-1}_2,x_3),(x_3,v^{t-1}_1),(v^{t-1}_1,v^{t-1}_2), \\ (v^{t-1}_1,u^{n-1}_2)$.
Recall that  $u^{n-1}c$ is an edge. Now $v^tu^{n-1}$ is not an
edge as otherwise $(x_3,v^t) \rightarrow (x_3,u^{n-1}) \in D$
while we had $(u^{n-1},x_3) \in D$ and we have an earlier circuit.
Now both $u^{n-1}_2,c$ are adjacent to $u^{n-1}$ and $v^t$ is not
adjacent to $u^{n-1}$ and $d,v^t$ both are adjacent to $g^{t-1}$
while $u^{n-1}_2$ is not adjacent to $g^{t-1}$. Therefore,
$(u^{n-1}_2,x_3)$ and $(x_3,v^t)$ are in the same component. \\

\noindent \textbf{Proof of 5:} Note that by (1) we have $(x_0,x_1)
\in S_1$. The proof of $(x_2,x_3) \in S_3$ is analogues to proof
of the (1) however for sake of completeness we give the proof.
Recall that $x_0p, x_1 q$ be the independent edges of $H$. Note
that $g^{t-1}x_2$ is not an edge as otherwise
$(v^{t-1}_2,g^{t-1})\rightarrow (v^{t-1}_2,x_2)$ and hence
we have an earlier circuit
$(x_2,x_3),(x_3,v^{t-1}_1),(v^{t-1}_1,v^{t-1}_2),(v^{t-1}_2,x_2)$.
Also $dg^{t-1}$ is an edge as otherwise $x_3d,g^{t-1}v^{t-1}$ are
independent edges and hence $(x_3,v^{t-1})$ would be in $D$, and
we obtain $(x_3,x_0)$ in less than $t$ steps. Now
$x_2x_0,dx_0,cx_0$ are edges of $H$ while $v^tx_0$ is not an edge
of $H$ and $dg^{t-1},cg^{t-1},v^tg^{t-1}$ are edges of $H$ while
$x_2g^{t-1}$ is not an edge and hence $(x_2,x_3),(x_3,v^t)$ are in
the same  component.

Now it remains to show $S_1=S_3$. Recall that $f^{m-1}w^m$ is an
edge of $H$, also $f^{m-1}a$ and $f^{m-1}q$ are edges of $H$ as
otherwise $f^{m-1}w^{m-1},x_1q$ are independent edges and
$f^{m-1}w^{m-1},x_1a$ are independent edges and hence
$(x_1,w^{m-1})$ is in  component and we obtain $(x_1,w^1)$ in less
than $m$ steps. $x_3d,ux_2$ are independent edges. Note that
$x_1u$ is not an edge as otherwise $(u,x_3)$ dominates $(x_1,x_3)$
and hence we get an earlier circuit. Recall that $x_2w^m$ is not
an edge. Moreover $v^{t}v^{t-2}$ is not an edge as otherwise
$x_3d,v^tv^{t-2}$ are independent edges and hence $(x_3,v^{t-2})$
is in a  component and we get $(x_3,v^1)$ in less than $t$ steps.

If $w^{m}v^t$ is an edge of $H$  then $(d,v^t) \rightarrow  (x_0,v^t) \rightarrow (x_0,w^m)$ and hence $S_1=S_3$.  So we may assume that $w^mv^t$ is not an edge. If $v^tq$ is an edge of $H$ then $(d,v^t) \rightarrow (x_0,v^t) \rightarrow (x_0,q)$ and hence $S_0=S_3$ and by (2) $S_0=S_1=S_2=S_3$.
If $w^md$ is an edge then $(a,w^m) \rightarrow (x_2,w^m) \rightarrow (x_2,d)$ and hence $S_1=S_3$. So we may assume $w^md$ is not an edge.

We conclude that $f^{m-1}g^{t-1}$ is an  edge as otherwise $(a,w^m) \rightarrow (d,w^m) \rightarrow (d,f^{m-1}) \rightarrow
(g^{t-1},f^{m-1}) \rightarrow  (g^{t-1},q) \rightarrow  (v^t,q)   \rightarrow (v^t,d)$, implying that $S_2= S_3'$ a contradiction.

Now $(a,w^m) \rightarrow (x_2,w^m) \rightarrow (x_2,f^{m-1}) \rightarrow (u, f^{m-1}) \rightarrow (u, g^{t-1}) \rightarrow (x_2, g^{t-1}) \rightarrow (x_2, d)$.
This would imply that $S_1=S_3$. \qed

\begin{lemma} \label{circuit3ap}
If we encounter a minimal circuit
$C=(x_0,x_1),(x_1,x_2),\dots,(x_3,x_0)$ at line 16 then there is a
component $S$ such that the envelope of every complete set $D_1$
where $S \subseteq D_1$ contains a circuit.
\end{lemma}
\pf We first consider the case that $S_i \ne S_j$ for some $i, \in
\{0,1,2,3\}$. According to Lemma \ref{relationship} we may assume
that $(x_0,x_1)$ is a simple pair in a component $S_0$ and
$(x_1,x_2)$ is a simple pair implied by component $S_1$, and none
of the $S_2$ and $S_3$ is in set $\{S_0,S_1\}$. In this case we
claim the following.


\begin{claim} \label{does_not_matter}
Suppose for some $1 \le i \le n$, $(u^i,u^i_1)$ is a simple pair
inside  component $R_1$ and $(u^i_1,u^i_2)$ is a simple pair
implied by a  component $R_2$.
 Then for any selection $R_3$ from $\{R_1,R'_1\}$ instead of $R_1$ and any selection $R_4$ from $\{R_2,R'_2\}$ instead of $R_2$ in lines 4--12;
 the pair $(u^{i-1}_2,x_3)$ is in $D$, and hence the complex pair $(x_2,x_3)$ is in $D$.
 \end{claim}
 \pf Note that since $u^iu^i_2$ is an edge, $(u^i_1,u^i_2)$ is implied by a  component.
Let $u^ia_i,u^i_1b_i$ be the independent edges and
$u^i_1c_i,d_ie_i$ be independent edges such that $(u^i_1,e_i)$ implies
$(u^i_1,u^i_2)$. Note that $u^i_2e_i$ and $u^i_2c_i,u^i_2b_i$ are
edges of $H$. Note that $(u^i,u^i_1)$ implies $(u^i,c_i)$ and
$(c_i,d_i)$ is in a  component. Thus $d_iu^i$ is not an edge as
otherwise $(c_i,d_i)$ dominates $(c_i,u^i)$ and we get a shorter
circuit. Similarly $a_ie_i$ is not an edge as otherwise
$(u^i_1,e_i)$ dominates $(u^i_1,a_i)$ a contradiction. Now
$e_iu^{i-1}_2$ is an edge as otherwise $u^iu^{i-1}_2,e_id_i$ are
independent edges and since $(u^i,e_i)$ is in $D$ (all the
 components have been added) , $(u^{i-1}_2,e_i)$
implies $(u^{i-1}_2,u^i_2)$ and we obtain $(u^1,x_3)$ in less than
$n$ steps. Also $u^{i-1}_2b_i,u^{i-1}_2c_i$ are edges of $H$ as
otherwise  $u^{i-1}_2u^i,u^i_1b_i$ or $u^{i-1}_2u^i,u^i_1c_i$
 are independent edges and hence $(u^{i-1}_2,u^i_1)$ is in a  component and we obtain $(u^1,x_3)$ in less than $n$ steps.

{\em Now this would imply that no matter what the algorithm selects from one of  $S_{u^iu^i_1},S_{u^i_1u^i}$ at lines 4--12
and no matter what the algorithm selects from one of the $S_{u^i_1e_i},S_{e_iu^i_1}$ at lines 4-12,
one of the pair $(u^i,u^i_2)$, $(e_i,u^i_2)$, and $(c_i,u^i_2)$ appears in $N^*[D]$. }

 Suppose we should have selected $S_{e_iu^i_1}$ and $S_{u^i_1u^i}$ lines 4--12. Now $(u^i_1,u^i)$ dominates
 $(u^i_1,u^i_2)$ and hence we have $(e_i,u^i_2)$. Thus $(e_i,x_3) \in D$ which implies $(u^{i-1}_1,x_3) \in D$.
 This means that instead of pair $(u^i,x_3)$ we would have $(e_i,x_3)$ and we would apply the same decomposition for $(e_i,x_3)$ as decomposition of $(u^i,x_3)$.
 If we should have selected $(c_i,d_i)$ and $(d_i,a_i)$ then $(d_i,u^i)\rightarrow(d_i,u^i_2)$ and hence $(c_i,u^i_2)$ would be in $D$ implying that $(c_i,x_3) \in D$ which
 would imply $(u^{i-1},x_3) \in  D$. The similar argument is implied for different selections of $R_3,R_4$.  \qed

\begin{claim} \label{doesnotmatter}
Suppose $(x_0,x_1)$ is a simple pair in component $S_0$ and
$(x_1,x_2)$ is a simple pair implied by  component $S_1$ such that
none of the $S_2$ and $S_3$ is in set $\{S_0,S_1\}$. Then by
replacing $S_0$ with $S'_0$ in $D$ or by replacing $S_{1}$ with
$S'_{1}$ in $D$ and keeping the  components $S_{2},S_{3}$ in $D$
 we still get a circuit
$(y_0,y_1),(y_1,y_2),(x_2,x_3),(x_3,y_0)$ in $N^*[D]$.
\end{claim}
\pf According to Claim \ref{does_not_matter} since we keep $S_2$
in $D$  the pair $(x_2,x_3)$ appears in $D$ (envelope
of $D$) . Since $(x_0,x_1)$ is a simple pair and
$x_0,x_1$ have different colors, there are independent edges
$x_0p,x_1q$. There are independent edges $x_1a,wb$ such that
$(x_1,w)$ implies $(x_1,x_2)$. Note that $x_2q,x_2x_0,x_2a$ are
edges since $(x_1,x_2)$ is not in a  component. As we argued
before in the correctness of lines 4--12,  $x_0b,pw$ are not edges of
$H$.  $qv$ is an edge as otherwise $(x_0,q) \rightarrow (v,q)
\rightarrow (v,x_2)$ and hence $(v,x_2) \in D$, yielding a shorter
(earlier) circuit $(x_2,x_3),(x_3,v),(v,x_2)$ which is a
contradiction. Suppose first both $qb,ap$ are edges of $H$. This
implies that $S_{x_0x_1}=S_{x_1w}$ and $(x_0,w) \in S_{x_0x_1}$.
We note that $wv$ is an edge as otherwise $(x_0,w) \rightarrow
(v,w) \rightarrow (v,x_2)$ and hence $(v,x_2) \in D$, yielding a
shorter (earlier) circuit $(x_2,x_3),(x_3,v),(v,x_2)$ which is a
contradiction. We conclude that $(x_3,v)$ implies $(x_3,w) \in D$.
Now if we choose $S'$ instead of $S_1$ at step (2) then we would
have $(x_1,x_0) \in D$ and $(x_1,x_0) \rightarrow (x_1,x_2) \in D$
and $(b,x_1) \in D$. Now we would have the circuit
$(w,x_1),(x_1,x_2),(x_2,x_3),(x_3,w)$. We now assume $qb$ is not
an edge. Proof for the case $ap \not\in E(H)$ is similar. $wv$ is
an edge as otherwise $(q,w) \rightarrow  (v,q) \rightarrow
(v,x_2)$, and again we get an earlier circuit. Now suppose we
would have chosen $(w,x_1)$ instead of $(x_1,w)$ in lines 4--12.
Note that $(w,p)$ is in a  component. Now either we have $S_{wp}
\in D$ or $S_{pw} \in D$. We continue by the first case $S_{wp}
\in D$ . We have $(w,p) \in D$ and $(p,q) \in D$ that
implies $(p,x_2)$ and hence we would have the circuit
$(w,p),(p,x_2),(x_2,x_3),(x_3,w)$. If $S_{pw} \in D$ in lines 4--12
then we have $(x_0,b) \in D$. Furthermore $(b,q)$ dominates
$(b,x_2)$ and now $(x_0,b),(b,x_2),(x_2,x_3),(x_3,x_0)$ would be a
circuit in $D$. By symmetry the other choices would yield a
circuit in $D$. \qed \\

%

\begin{figure}[htbp] 
 \begin{center}
 \includegraphics[scale=0.65]{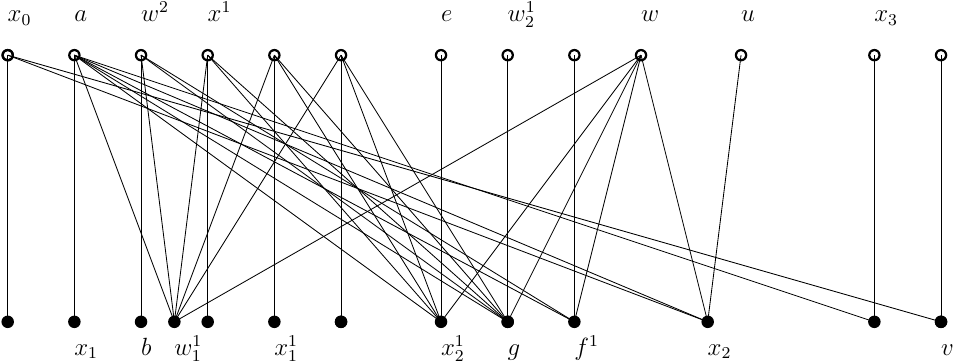}
 \caption{Another example for decomposition of  $(x_1,w)$. We have components $S_{x_1w^2}=S_{x^1_2g}=S_{gf^1}$.}
 \label{fig:fig6}
 \end{center}
 \end{figure}

{\textbf {Remark : }} The decomposition was for each of the pair $(x_0,x_1),(x_1,x_2),(x_2,x_3), (x_3,x_0)$.
Now for example consider the complex pair $(x_1,x_2)$ implied by $(x_1,w)$. When we decompose $(x_1,w)$ into
pairs $(x_1,w^1_1),(w^1_1,w^1_2),(w^1_2,w)$ then we
recursively decompose $(x_1,w^1_1)$.
By applying the decomposition to each of the $(w^1_1,w^1_2),(w^1_2,w)$ we
reach the same conclusion as for the pairs $(x_0,x_1),(x_1,x_2),  (x_2,x_3),(x_3,x_0)$.
In fact the circuit $C$ has four pairs that we can view as external pairs while
the pair $(w^1_2,w)$ is an internal pair and the same rule applied for it with respect to pair $(w^1_1,w^1_2)$  
(see Figure \ref{fig:fig6}). \qed

\begin{lemma}\label{dictator_correctly_ap}
The algorithm computes the $DCT(x,y)$ correctly.
\end{lemma}
\pf  Suppose by adding pair $(x,y)$ into $D$ we close a circuit ,and there was no circuit before that. By Corollary \ref{circuit-lengthap}, a minimal circuit
$C$ has four vertices and we may assume $C=(x_0,x_1),(x_1,x_2),(x_2,x_3), \\ (x_3,x_0)$.
W.l.o.g assume that $x_0,x_3$ are white vertices and $x_1,x_2$ are black vertices.

Recall that the followings determine the dictatorship of a pair $(x,y)$.

 \begin{itemize}
\item [(a)] If $(x,y) \in N^+[S]$ for some  component $S$ then
$DCT(x,y)=S$. \item [(b)] If $x,y$ have different colors and
$(x,y)$ is implied by some pair $(u,y)$ then
$DCT(x,y)=DCT(u,y)$. \item [(c)] If $x,y$ have the same color
and $(x,y)$ is implied by some pair $(x,w)$ then
$DCT(x,y)=DCT(x,w)$. \item [(d)] If $x,y$ have the same color
and $(x,y)$ is by transitivity on $(x,w),(w,y)$ then
$DCT(x,y)=DCT(w,y)$. \item [(e)] If $x,y$ have different colors
and $(x,y)$ is by transitivity on $(x,w),(w,y)$ then
$DCT(x,y)=DCT(x,w)$.
\end{itemize}

Suppose $(u,x_3) \in N^*[D]$ and $(u,x_3)  \rightarrow (x_2,x_3)$. 
According to definition $DCT(u,x_3)=DCT(x_2,x_3)$. Since $(u,x_3)$ is by transitivity, by Corollary \ref{chain1ap} 
we have the pairs $(u,u_1),(u_1,u_2),(u_2,x_3)$ in $N^*[D]$.
When we compute $N^*[D]$, $(u,x_3)$ appears in $N^*[D]$ whenever $(u,f)$ and $(f,x_3)$ appeared in $N^*[D]$ at some earlier level.
According to minimality of the chain $(u,x_3)$ either $f=u_2$ or $f=u_1$. First suppose $f=u_2$. Now according to (d)  we have
$DCT(x_2,x_3)=DCT(u_2,x_3)$. By induction hypothesis we know that $DCT(u_2,x_3)=S_2$. Recall that $S_2$ is the component obtained after decomposing of $(u,x_3)$ in Lemma \ref{circuit3ap}. Therefore, $DCT(x_2,x_3)=DCT(u,x_3)=DCT(u_2,x_3)$. Now consider the case $f=u_1$.
According to (d) we have $DCT(u,x_3)=DCT(u_1,x_3)$. In this case by using (e) we have $DCT(u_1,x_3) =DCT(u_2,x_3)$
because $(u_1,u_2),(u_2,x_3)$ imply $(u_1,x_3)$ and $u_1,x_3$ have different colors.
Similar argument is implied for pair $(x_1,x_2)$, where $x_1,x_2$ have the same color. \qed

\section{Correctness of lines  13--21}

If we encounter a circuit $C$ in $D$ after line 12 then according to
Lemma \ref{circuit3ap} there is a  component $S$ that is a
dictator for $C$. We compute this dictator component by $DCT$
function (also by decomposing the pairs of the circuit as
explained in Section \ref{circuit-structure} ) and its correctness is justified by Lemma
\ref{dictator_correctly_ap}.

It is clear that we should not add $S$ to $D$ as otherwise we won't be able to obtain the desired ordering. 
Therefore, we must take the coupled component of every dictator component of a circuit appeared at the first time we take the envelope of $D$. Now we continue to show the correctness of 13--21. 

\begin{lemma}\label{responsible-noteap}
If all the  components $S_{ab},S_{ba},S_{bc},S_{cb},S_{ac},S_{ca}$
are pairwise distinct then none of them is a dictator component.
\end{lemma}
\pf By the assumption of the lemma, $H$ is pre-insect with $Z=
\emptyset$. Now as we argued in Section \ref{circuit-structure}, if component
$S$ is a dictator for a circuit then there has to be pairs
$(x,y),(y,z),(x,z) \in S$. However according to the structure of
pre-insect $S_{ab}$ consists of only the pairs $(x,y)$ that $x \in
H_1$ and $y \in H_2$. \qed

\begin{lemma}\label{responsibleap}
If for every $S \in \mathcal{DT}$ we add $N^+[S']$ into $D_1$ and for every $R \in D \setminus \mathcal{DT}$ we add $N^+[R]$ into $D_1$ at line 18 then we do not encounter a circuit 
in $N^+[D_1]$. 
\end{lemma}
\pf Suppose we encounter a shortest circuit $(x_0,x_1),(x_1,x_2),...,(x_{n-1},x_n),(x_n,x_0)$ with the simple pairs such that
at least one pair $(x_i,x_{i+1})$ belongs to some $N^+[S']$, $S \in \mathcal{DT}$ ( $\mathcal{DT}$ is the set of the dictator components).

We say $(x_i,x_{i+1})$ is an old pair if it is in $N^+[S]$ and $S \not\in
\mathcal{DT}$. Otherwise $(x_i,x_{i+1})$ is called a new pair.
First suppose that both $(x_i,x_{i+1}),(x_{i+1},x_{i+2})$ are in
components. By Corollary \ref{jing3.2(C)ap} $(x_i,x_{i+2})$ is
also in a component. Now if both $(x_i,x_{i+1})$,
$(x_{i+1},x_{i+2})$ are old then $(x_i,x_{i+2})$ is
also an old pair. Otherwise we have $S_{x_ix_{i+1}} \ne
S_{x_ix_{i+2}}$, $S_{x_{i+1}x_{i+2}} \ne S_{x_ix_{i+2}}$ and
$S_{x_ix_{i+1}} \ne S_{x_{i+1}x_{i+2}}$, moreover $S_{x_ix_{i+1}}
\ne S_{x_{i+2}x_{i+1}}$, $S_{x_{i+1}x_{i}} \ne S_{x_{i+1}x_{i+2}}$
because there was no circuit at lines 4--12. Now $H$ is a pre-insect
with $Z= \emptyset$ and hence by Lemma \ref{responsible-noteap},
$S_{x_ix_{i+2}}$ is not a dictator component. Similarly according
to the minimality of the circuit, it is not possible that both
$(x_i,x_{i+1})$ and $(x_{i+1},x_{i+2})$ are new. So we may assume
that $(x_i,x_{i+1})$ is old and $(x_{i+1},x_{i+2})$ is new. Now
again we know that $S_{x_ix_{i+1}} \ne S_{x_{i+1}x_{i+2}}$ and
$S_{x_ix_{i+1}} \ne S_{x_ix_{i+2}}$. We note that
$S_{x_{i+1}x_{i+2}} \ne S_{x_ix_{i+2}}$ as otherwise we get a
shorter circuit. Therefore, $H$ is pre-insect with $Z= \emptyset$
and hence by Lemma \ref{responsible-noteap} $(x_{i+1},x_{i+2})$ is
not in a dictator component.

If none of the $(x_i,x_{i+1})$, $(x_{i+1},x_{i+2})$ is in a
component, then $(x_i,x_{i+2})$ is implied by the same component
implying $(x_i,x_{i+1})$ and hence we get a shorter circuit. So we
may assume that $(x_i,x_{i+1})$'s alternate, meaning that if
$(x_i,x_{i+1})$ is implied by a  component then
$(x_{i+1},x_{i+2})$ is in a  component and vice versa. Now in this
case as we argue in the correctness of lines 4--12 there would be an
exobiclique in $H$ which is not possible. \qed \\

We present the following lemma as a remark on the number of distinct dictator components. 

\begin{lemma}\label{number}
The number of distinct dictator  components is at most $2n$.
\end{lemma}
\pf Note that there are at most $n^2$ distinct  components.
Consider  component $S_{ab},S_{ac}$ such that $S_{ab} \ne S_{ac}$
and $S_{ab} \ne S_{ca}$. It is not difficult to see that $S_{bc}$
is also a  component as otherwise $S_{ab}=S_{ac}$. Now we must
have $S_{bc}=S_{ac}$ or $S_{bc}=S_{ca}$ as otherwise by Lemma
\ref{responsible-noteap}, $S_{ab}$ would not be a dictator
component. In general, if vertex $a$ with vertices
$a_1,a_2,...,a_k$ appear in distinct dictator components
$S_{aa_i}$, $1 \le i \le k$ then none of the $S_{a_ia_j}$ would be
distinct from $S_{aa_1},S_{aa_2},...,S_{aa_k}$. These would imply
that there are at most $O(n)$ distinct dictator components. \qed

\section{Correctness of the lines 22--23 }
\begin{tm}\label{heightap}
By always choosing a component $S \in H^+ \setminus D$ with $N^+[S]=S$, and taking transitive closure,
the algorithm does not create a circuit.
\end{tm}
\pf Suppose by adding a terminal (trivial) component $(x,y)$ into $D$ we create a circuit. Note that none of the $(x,y),(y,x)$ is in $D$ and also $(x,y)$
is not by transitivity on some of the pairs in $D$ as otherwise it would be placed in $D$. Since $(x,y)$ is a sink pair at the current step of the algorithm,
if $(x,y)$ dominates a pair $(u,v)$ in $H^+$ then $(u,v)$ is in $D$. The only way that adding $(x,y)$ into $D$ creates a circuit in $D$ is when $(x,y)$
dominates a pair $(u,v)$ while there is a chain $(v,y_1),(y_1,y_2),...,(y_k,v)$ of pairs in $D$ implying that $(v,u) \in D$.
When $x,y$ have the same color $v=y$ and $xu$ is an edge which means $(v,u)$ implies $(y,x)$ and hence $(y,x) \in D$ a contradiction.
When $x,y$ have different colors then $u=x$ and $yv$ is an edge and hence $(v,x) \in D$ where $(v,x) \rightarrow (y,x)$ a contradiction. \qed

\section{Implementation and complexity }

In order to construct digraph $H^+$, we need to list all the
neighbors of each vertex. If $x,y$ in $H$ have different colors
then pair $(x,y)$ of $H^+$, has $d_y$ out-neighbors where $d_y$
is the  degree of $y$ in $H$. If $x,y$ have the same color then
pair $(x,y)$ has $d_x$ out-neighbors in $H^+$. For simplicity we
assume that $|W|=|B|=n$. For a fixed black vertex $x$ the number
of all pairs where each of them is  a neighbor of all pairs $(x,z)$,  $z
\in V(H)$, is  $nd_x+d_{y_1}+d_{y_2}+\dots +d_{y_n}$,
$y_1,y_2,...,y_n$ are all the white vertices. Therefore, it takes
$O(nm)$, $m$ is the number of edges in $H$, to construct $H^+$. We
may use a linked list structure to represent $H^+$. It order to
check whether there exists a self-coupled component, it is enough
to see whether $(a,b)$ and $(b,a)$ belongs to the same component.
This can be done in time $O(mn)$ using Tarjan's strongly connected component algorithm. Since we maintain a partial
order $D$ once we add a new pair into $D$ we can decide whether we
close a circuit or not. Computing $N^*[D]$ takes $O(n(n+m))$
since there are $O(mn)$ edges in $H^+$ and there are at most
$O(n^2)$ vertices in $H^+$. Note that the algorithm computes the
envelope of $D$ at most twice; once at line 15, and once at 
line 25. 

Once a pair $(x,y)$ is added into $D$, we put an arc from $x$ to $y$ in the partial order and the arc $xy$ gets a time label denoted by $T(x,y)$.
$T(x,y)$ is the level in which $(x,y)$ is added. In order to look for a circuit we need to consider a circuit $D$ in which each pair is original.
Once a circuit is formed at lines 13--21;  we can find a dictator component $S$ by using $DCT$ function, and store $S$ into set $\mathcal{DT}$.
Therefore, we spend at most $O(nm)$ time to find all the dictator components. At lines 24-25, we add the rest of the remaining pairs and that takes at most $O(n^2)$. 
Now it is clear that the running time of the algorithms is $O(nm)$. \\

{\textbf {Acknowledgment :}} The author would like to thank Pavol Hell and Jing Huang for many valuable discussions and for many helps in the early stage of this paper.


\begin{thebibliography}{40}

\bibitem{benzer} S. Benzer.  On the topology of the genetic fine structure. {\em Proc. Natl. Acad. Sci. USA}  45 : 1607--1620 (1959).

\bibitem{booth} K.S. Booth and  G.S. Lueker.  Testing for the consecutive ones property, interval graphs, and graph planarity using PQ-tree algorithms.  
{\em J. Comput. System Sci} 13 (3):  335--379 (1976).

\bibitem{bls} A. Brandst\"adt, V. B. Le, and J. P. Spinrad.
{\textbf {Graph Classes : SIAM Monographs Discrete Math and Applications}}, Philadelphia, 1999.


\bibitem{denver} D. E. Brown, J. R. Lundgren, and S. C. Flink.
 Characterizations of interval bigraphs and unit interval bigraphs.
{\em Congressus Num}, 157 : 79 -- 93 (2002) 

\bibitem{corneil}
D.G. Corneil, S.Olariu, and L. Stewart.  The ultimate interval graph recognition algorithm.  
{\em SODA},  175--180 (1998).

\bibitem{corneil09} D.G. Corneil, S.Olariu, L.Stewart. The LBFS Structure and Recognition of Interval Graphs. {\em  SIAM J. Discrete Math.}, 23(4) : 1905--1953 (2009) .

\bibitem{dama} P. Damaschke. Forbidden Ordered Subgraphs. {\em Topics in Combinatorics and Graph Theory}, 219--229 (1990).

\bibitem{dgr}  D. Duffus, M. Ginn, V. R\"{o}dl. On the computational complexity of ordered subgraph recognition. {\em Random Structures and Algorithms}, 
7 (3) : 223--268 (1995).

\bibitem{adjust} T. Feder, P. Hell, J.Huang and A. Rafiey.  Interval Graphs, Adjusted Interval Digraphs, and Reflexive
List Homomorphisms. {\em Discrete Applied Math},  160(6) : 697--707 (2012).

\bibitem{golumbic} M. C. Golumbic. {\em Algorithmic Graph Theory and Perfect Graphs, 2nd ed., Ann. Discrete Math.
57}, Elsevier, Amsterdam, The Netherlands, 2004.

\bibitem{habib}
M.Habib, R.McConnell, Ch.Paul and L.Viennot. 
 Lex-BFS and partition refinement, with applications to transitive orientation, interval graph recognition, and consecutive ones testing. 
 {\em Theor. Comput. Sci.}, 234 : 59--84 (2000).

\bibitem{hkm} F. Harary, J.A. Kabell and F.R. McMorris.  Bipartite intersection
graphs. {\em Comment, Math Universitatis Carolinae}, 23 : 739 -- 745 (1982).

\bibitem{hh2003} P. Hell and J. Huang.  Interval bigraphs and circular arc graphs. 
{\em J. Graph Theory}, 46 : 313 -- 327 (2003) .


\bibitem{esa-2012} P.Hell, M.Matrolilli, M.Nevisi and A.Rafiey.  Approximation of Minimum Cost Homomorphisms. {\em ESA }, 587--598 (2012).  


\bibitem{esa-2014} P.Hell, B.Mohar and A.Rafiey. Ordering without Forbidden Patterns. {\em ESA },  554--565 (2014)


\bibitem{korte} N. Korte, Rolf H. M\"ohring. An Incremental Linear-Time Algorithm for Recognizing Interval Graphs. {\em SIAM J. Comput.}, 18(1) : 68--81 (1989).

\bibitem{ross} R.M. McConnell. Linear time recognition of circular-arc graphs. {\em IEEE FOCS }, 42 : 386--394 (2001) .

\bibitem{muller} H.M\"uller.  Recognizing interval digraphs and interval bigraphs
in polynomial time. {\em Discrete Appl. Math.}, 78 : 189 -- 205  (1997).

\bibitem{sdrw} M. Sen. S. Das, B. Roy, and D.B. West.  Interval digraphs:
An analogue of interval graphs. {\em J. Graph Theory}, 13 : 189 -- 202 (1989). 

\bibitem{bss} J.P.Spinrad, A. Brandst\"adt and L.Stewart. Bipartite permutation graphs. {\em Discrete Applied Mathematics}, 18 : 279--292 (1987).


\end{thebibliography}
\end{document}